\documentclass{article}

\usepackage{microtype}
\usepackage{booktabs} 

\usepackage{subcaption}
\usepackage{color,soul}
\usepackage{verbatim}
\usepackage{hyperref}

\usepackage[accepted]{icml2023}

\usepackage{amsmath}
\usepackage{amssymb}
\usepackage{mathtools}
\usepackage{amsthm}

\usepackage[capitalize,noabbrev]{cleveref}

\theoremstyle{plain}
\newtheorem{theorem}{Theorem}[section]

\newtheorem{lemma}[theorem]{Lemma}

\theoremstyle{definition}
\newtheorem{definition}[theorem]{Definition}

\theoremstyle{remark}

\usepackage[textsize=tiny]{todonotes}

\icmltitlerunning{Column Subset Selection and Nyström Approximation via Continuous Optimization}

\renewcommand{\v}[1]{\boldsymbol{#1}}
\newcommand{\m}[1]{\mathbf{#1}}

\newcommand{\bb}[1]{\mathbb{#1}}
\renewcommand{\c}[1]{\mathcal{#1}}

\newcommand{\tr}{\operatorname{tr}}

\DeclareMathOperator*{\argmin}{argmin}

\def\boxit#1{\vbox{\hrule\hbox{\vrule\kern6pt
          \vbox{\kern6pt#1\kern6pt}\kern6pt\vrule}\hrule}}

\begin{document}

\twocolumn[
\icmltitle{Column Subset Selection and Nyström Approximation via Continuous Optimization}



\icmlsetsymbol{equal}{*}

\begin{icmlauthorlist}
\icmlauthor{Anant Mathur}{yyy}
\icmlauthor{Sarat Moka}{yyy}
\icmlauthor{Zdravko Botev}{yyy}

\end{icmlauthorlist}

\icmlaffiliation{yyy}{Department of XXX, University of YYY, Location, Country}

\icmlcorrespondingauthor{Firstname1 Lastname1}{first1.last1@xxx.edu}
\icmlcorrespondingauthor{Firstname2 Lastname2}{first2.last2@www.uk}


\vskip 0.3in
]




\begin{abstract}
We propose a continuous optimization algorithm for the Column Subset Selection Problem (CSSP) and Nyström approximation. The CSSP and Nyström method construct low-rank approximations of matrices based on a predetermined subset of columns. It is well known that choosing the best column subset of size $k$ is a difficult combinatorial problem. In this work, we show how one can approximate the optimal solution by defining a penalized continuous loss function which is minimized via stochastic gradient descent. We show that the gradients of this loss function can be estimated efficiently using matrix-vector products with a data matrix $\m X$ in the case of the CSSP or a kernel matrix $\m K$ in the case of the Nyström approximation. We provide numerical results for a number of real datasets showing that this continuous optimization is competitive against existing methods.
\end{abstract}

\section{Introduction}

Recent advances in the technological ability to capture and collect data have meant that high-dimensional datasets are now ubiquitous in the fields of engineering, economics, finance, biology, and health sciences to name a few. In the case where the data collected is not labeled it is often desirable to obtain an accurate low-rank approximation for the data which is relatively low-cost to obtain and memory efficient. Such an approximation is useful to speed up downstream matrix computations that are often required in large-scale learning algorithms. The Column Subset Selection Problem (CSSP) and Nyström method are two such tools that generate low-rank approximations based on a subset of data instances or features from the dataset. The chosen subset of instances or features are commonly referred to as ``landmark" points. The choice of landmark points determines how accurate the low-rank approximation is.


The challenge in the CSSP is to select the best $k$ columns of a data matrix $\m X\in \bb{R}^{m\times n}$ that span its column space. That is, for any binary vector $\v s \in \{0, 1\}^n$, compute
\begin{equation}
\label{eq: comb_problem}
\argmin_{\m s \in \{0, 1\}^n} \| \m X - \m P_s\m X\|^2_{F},\quad\text{subject to }\|\v s\|_0 \leq  k,
\end{equation}
where $\|\cdot\|_F$ is the Frobenius matrix norm, $\|\v s\|_0 = \sum_{j=1}^n I(s_j=1)$ and $\m P_s$ is the projection matrix onto span$\{\v x_j: s_j =1, j=1,\ldots,n\}$ ($\v x_j$ being the $j$-th column of $\m X$). 

Solving this combinatorial problem exactly is known to be NP-complete \citep{shitov2021column}, and is practically infeasible even when $k$ is of moderate size. We propose a novel continuous optimization algorithm to approximate the exact solution to this problem. While an optimization approach via Group Lasso \citep{yuan2006model} exists for the convex relaxation of this problem \citep{bien2010cur}, to the best of our knowledge, no continuous optimization method has been developed to solve the highly non-convex combinatorial problem \eqref{eq: comb_problem}. To introduce our approach for the CSSP, instead of searching over binary vectors $\v s \in \{ 0,1\}^n$, we consider the hyper-cube $[0,1]^n$ and define for each $\v t \in [0,1]^n$ a matrix $\widetilde{\m P}(\v t)$ which allows the following well-defined penalized continuous extension of the exact problem,
\[\argmin_{\v t \in [0, 1]^n} \|\m X - \widetilde{\m P}(\v t)\m X\|_F^2 + \lambda\sum_{j = 1}^n t_j.\]
The parameter $\lambda>0$ plays an analogous role to that of the regularization parameter in regularized linear regression methods  \citep{tibshirani1996regression} and controls the sparsity of the solution, that is, the size of $k$. Two aspects of this continuous extension make it useful for approximating the exact solution. Firstly, the continuous loss agrees with the discrete loss at every corner point $\v s\in \{0,1\}^{n}$ of the hypercube $[0, 1]^n$, and secondly, for large datasets the gradient can be estimated via an unbiased stochastic estimate. To obtain an approximate solution to the exact problem, \emph{stochastic gradient descent} (SGD) is implemented on the penalized loss. After starting at an interior point of the hyper-cube, under SGD, the vector $\v t$ moves towards a corner point, and some of the $\v t_j$'s exhibit shrinkage to zero. It is these values that indicate which columns in $\m X$ should not be selected as landmark points.

The Nyström approximation \citep{williams2000using,drineas2005nystrom} is a popular variant of the CSSP for positive semi-definite kernel matrices. The Nyström method also constructs a low-rank approximation $\widehat{\m K}\in\bb{R}^{n\times n}$ to the true kernel matrix $\m K\in \bb{R}^{n\times n}$ using a subset of columns. Once the $k$ columns are selected, $\widehat{\m K}$ (in factored form) takes $O(k^3)$ additional time to compute, requires $O(nk)$ space to store, and can be manipulated quickly in downstream applications, e.g., inverting $\widehat{\m K}$ takes $O(nk^2)$ time. In addition to the continuous extension for the CSSP, in this paper, we provide a continuous optimization algorithm that can approximate the best $k$ columns to be used to construct $\widehat{\m K}$ (\cref{sec: nystrom}).

The continuous algorithm for the CSSP formulated in this paper utilizes SGD where at each iteration one can estimate the gradient with a cost of $O(mn)$. We show that the gradients of the penalized continuous loss can be estimated via linear solves with random vectors that are approximated with the conjugate gradient algorithm (CG) \citep{golub1996matrix}, which itself is an iterative algorithm that only requires matrix-vector multiplications (MVMs) with the $m\times n$ matrix $\m X$.  Similarly, for the Nyström method we show that at each step of the gradient descent, the gradient can be estimated in  $O(n^2)$ time requiring only matrix-vector multiplications with the kernel matrix  $\m K$. This is especially useful in cases where we only have access to a black-box MVM function. The fact that both these algorithms require only matrix-vector multiplications to estimate the gradients lends itself to utilizing GPU hardware acceleration. Moreover, the computations in the proposed algorithm can exploit the sparsity that is achieved by working only with the
 columns of $\m X$ that are selected by the algorithm at any given iteration.  

\subsection{Related Work}

There exists extensive literature on random sampling methods for the approximation of the exact CSSP and Nyström problem. Sampling techniques such as adaptive sampling \citep{deshpande2006adaptive}, ridge leverage scores \citep{gittens2013revisiting, musco2017recursive,alaoui2015fast} attempt to sample ``important" and ``diverse" columns. In particular, recent attention has been paid to Determinantal Point Processes (DPPs) \citep{hough2006determinantal,derezinski2021determinantal}. DPPs  provide strong theoretical guarantees \citep{derezinski2020improved} for the CSSP and Nyström approximation and are amenable to efficient numerical implementation \citep{li2016fast, derezinski2019exact,calandriello2020sampling,derezinski2019fast}. Outside of sampling methods, iterative methods such as Greedy selection \citep{farahat2011novel, farahat2013efficient} have been shown to perform well in practice and exhibit provable guarantees \citep{altschuler2016greedy}.

Column selection has been extensively studied in the supervised context of linear regression (more commonly referred to as feature or variable selection). Penalized regression methods such as the Lasso \citep{tibshirani1996regression} have been widely applied to select columns of a predictor matrix that best explain a response vector.  The canonical $k$-best subset or $l_0$-penalized regression problem is another penalized regression method, where the goal is to find the best subset of $k$ predictors that best fit a response $\v y$ \citep{beale1967discarding,hocking1967selection}. The recently proposed \emph{Continuous Optimization Method Towards Best Subset Selection} (COMBSS) algorithm \citep{moka2022combss} attempts to solve the $l_0$-penalized regression problem by minimizing a continuous loss that approximates the exact solution. The algorithm we propose for the CSSP in this paper can be viewed as an adaptation of COMBSS to the unsupervised setting. In this setting, the goal is to find the best subset of size $k$ for a multiple multivariate regression model where both the response and predictor matrix are $\m X$. Interestingly, this framework can be extended to include a continuous selection loss for the Nyström approximation. 

The rest of the paper is structured as follows.  In \cref{sec: cont ext} we  describe the continuous extension for the CSSP and the Nyström method. In \cref{sec: imp} we provide steps for the efficient implementation of our proposed continuous algorithm on large matrices and in \cref{sec: emp_eval} we provide numerical results on a variety of real datasets.

\section{Continuous Loss for Landmark Selection}
\label{sec: cont ext}
In this section, we formally define the CSSP and the best size $k$-Nyström approximation. Then, we provide the mathematical setup for the continuous extension of the exact problem.

\subsection{Column Subset Selection}
Let $\m X \in \bb{R}^{m \times n}$ and for any binary vector $\v s = (s_1,\dots,s_n)^{\top}\in \{0, 1\}^n$, let $\m X_{[\v s]}$ denote the matrix of size $m\times \|s\|_0$ keeping only columns $j$ of $\m X $ where $s_j =1$, for $j = 1,\dots, n$. Then for every integer $k\leq n$ the CSSP finds
\begin{equation}
\label{bcss}
    \argmin_{\m s \in \{0, 1\}^n} \| \m X - \m P_s\m X\|^2_{F},\quad\text{subject to }\|\v s\|_0 \leq  k, 
\end{equation}
where $\m P_s := \m X_{[\v s]}\m X_{[\v s]}^{\dagger}$ ($\dagger$ denotes Moore–Penrose inverse) is the projection matrix onto $\textrm{span}\{\m x_j: s_j=1\}$ and $\v x_j$ is the $j$-th column of $\m X$. By expanding the Frobenius norm it is easy to see that the discrete problem \eqref{bcss} can be re-formulated as,
\begin{equation*}
\label{bcss2}
\argmin_{\m s \in \{0, 1\}^n} -\tr\left[\m X^{\top}\m P_{\v s}\m X\right],\quad\textrm{subject to }\|\v s\|_0 \leq k.
\end{equation*}

We now define a new matrix function on $\v t \in[0,1]^n$ which acts as a continuous generalization of $\m P_{\v s}$.
\begin{definition}
\label{def:p(t)}
For $\v t = (t_1,\dots,t_n)^{\top}\in [0,1]^n$, define $\m T := \operatorname{Diag}(\v t)$ as the diagonal matrix with diagonal elements $t_1,\dots,t_n$ and
\begin{equation*}
\label{Pt}
    \widetilde{\m P}(\v t) := \m X\m T \left[\m T\m X^{\top}\m X\m T+\delta(\m I-\m T^2)\right]^{\dagger}\m T\m X^\top,
\end{equation*}
where $\delta>0$ is a fixed constant. 
\end{definition}

Although not explicitly stated in \citep{moka2022combss}, $\widetilde{\m P}(\v t)$ is used as the continuous generalization for the hat matrix $\m P_{\v s}$ to solve the $l_0$-penalized regression problem. 

The main difference between this definition and traditional sampling methods is that instead of multiplying $\m X$ by a sampling matrix to obtain $\m X_{[\v s]}$ we compute the matrix $\m X \m T$ which weights column $j$ of $\m X$ by the parameter $t_j\in[0,1]$. Intuitively, the matrix $\m T\m X^{\top}\m X\m T+\delta(\m I-\m T^2)$ can be viewed as a convex combination of the matrices $\m X^{\top}\m X$ and $\delta \m I$. 

From an evaluation standpoint, the pseudo-inverse need not be evaluated for any interior point in this newly defined function.
We remark that for any $\v t \in [0,1)^n$ the matrix inverse in \cref{def:p(t)} exists and therefore, 
\[
\widetilde{\m P}(\v t) = \m X\m T \left[\m T\m X^{\top}\m X\m T+\delta(\m I-\m T^2)\right]^{-1}\m T\m X^\top.
\]
We now state two results for the function  $\widetilde{\m P}(\v t)$ and its relationship with the projection matrix $\m P_s$. The following Lemmas (\ref{lem: cnr_p} and \ref{lem: cont_p}) are extensions of the results stated in \citep{moka2022combss}.
\begin{lemma}
\label{lem: cnr_p}
For any binary vector $\v s\in\{0,1\}^n$, $\widetilde{\m P}(\v s)$ exists and
\[\widetilde{\m P}(\v s) = \m P_{\v s} =  \m X_{[\v s]}  \m X_{[\v s]}^{\dagger}.\]
\end{lemma}

\begin{lemma}
\label{lem: cont_p}
$\widetilde{\m P}(\v t)$ is continuous element-wise over $[0,1]^n$. Moreover, for any sequence $\v t^{(1)},\v t^{(2)}\dots\in[0,1)^n$ converging to $\v t\in[0,1]^n$, the limit $\lim_{l\to\infty} \widetilde{\m P}(\v t^{(l)})$ exists and
\[\lim_{l\to\infty} \widetilde{\m P}(\v t^{(l)}) = \widetilde{\m P}(\v t).\]
\end{lemma}
We note that the proof of \cref{lem: cont_p} follows identically to the proof of \emph{Theorem 3} in \citep{moka2022combss} where it is stated that the function $\|\v y - \widetilde{\m P}(\v t)\v y\|_2^2$ is continuous over $[0, 1]^{n}$ for any fixed vector $\v y \in \bb{R}^{n}$. 

Given $\widetilde{\m P}(\v t)$ is continuous on $[0,1]^n$ and agrees with $\m P_{\v s}$ at every corner point we can define the continuous generalization of the exact problem 
\eqref{bcss},
\begin{equation*}
    \argmin_{\v t \in [0, 1]^n} -\tr\left[\m X^{\top}\widetilde{\m P}(\v t)\m X\right],\quad\text{subject to }\sum_{j=1}^n t_j \leq  k.
\end{equation*}
Instead of solving this constrained problem, for a tunable parameter $\lambda$, we consider minimizing the Lagrangian function,
\begin{equation*}
\label{eq: combss}
     \argmin_{\v t \in [0, 1]^n}  f_{\lambda}(\v t),\quad f_{\lambda}(\v t):=-\tr\left[\m X^{\top}\widetilde{\m P}(\v t)\m X\right]+\lambda\sum_{j=1}^n t_j.
\end{equation*}
In \cref{sec: imp} we reformulate this box-constrained problem into an equivalent unconstrained problem via a nonlinear mapping $\v t = \v t(\v w)$ for $\v w \in \bb{R}^n$ that forces $\v t$ to be in the hypercube $[0,1]^n$. We solve this optimization via continuous gradient descent. To this end, we need to evaluate the gradient $\nabla f_{\lambda}(\v t)$ for any interior point.

\begin{lemma}
\label{lemma: f_grad}
Let $\m K = \m X^{\top}\m X$, $\m Z =\m K - \delta \m I$ and $\m L_{\v t} = \m T \m Z \m T +\v\delta\m I$. Then, for $\v t \in (0,1)^n$,
\begin{equation*}
    \nabla f_{\lambda}(\v t) = 2\operatorname{Diag}\left[\m L_{\v t}^{-1}\m T \m K^2 \left(\m T\m L_{\v t}^{-1}\m T\m Z - \m I  \right)\right]+\lambda\v 1.
\end{equation*}
\end{lemma}
Evaluating $\nabla f_{\lambda}(\v t)$ has a computational complexity of $O(n^3)$ due to the required inversion of $\m L_t$. In \cref{sec: imp} we detail an unbiased estimate for $\nabla f_{\lambda}(\v t)$ which utilizes the CG algorithm, where the most expensive operations involved are matrix-vector multiplications with $\m X$ and $\m X^{\top}$, which reduces the computational complexity to $O(mn)$.

\subsection{Nyström Method}
\label{sec: nystrom}
We now turn our attention to defining a continuous objective for the landmark points in the Nyström approximation. We consider optimizing the landmark points first with respect to the trace matrix norm and then to the Frobenius matrix norm.

In many applications, we are interested in obtaining a low-rank approximation to a \emph{kernel} matrix $\m K \in \bb{R}^{n\times n}$. Consider an input space $\c X$ and a positive semi-definite kernel function $h :\c  X \times \c X \rightarrow \bb R$. Given a set of $n$ input points $\v x'_1, . . . , \v x'_n \in \c X $,
the kernel matrix $\m K \in \bb{R}^{n\times n}$ is defined by $\m K_{i,j} = h(\v x'_i, \v x'_j )$ and is positive semi-definite. 

For any binary vector $\v s \in \{0, 1\}^n$ let $\m K_{[\v s]}$ be the $n\times\|\v s \|_0$ matrix with columns indexed by $\{j: s_j = 1\}$ and $\m K_{[\v s,\v s]}$ be the $\|\v s \|_0\times\|\v s \|_0$ principal sub-matrix indexed by $\{j: s_j = 1\}$. The Nyström low-rank approximation for $\m K$ is given by,
\begin{equation*}
    \widehat{\m K}_s := \m K_{[\v s]}\m K_{[\v s,\v s]}^{\dagger}\m K_{[\v s]}^{\top}.
\end{equation*}
The following observation appearing in \citep{derezinski2020improved} connects the CSSP and the Nyström approximation with respect to the trace matrix norm.

Suppose we have the decomposition of the kernel matrix  $\m K = \m X^{\top}\m X$ where $\m X\in\bb{R}^{m \times n}$. Then, the Nyström approximation is given by $\widehat{\m K}_{\v s} = \left(\m P_{\v s} \m X\right)^{\top}\m P_{\v s}\m X$ and
\[
 \|\m K - \widehat{\m K}_{\v s}\|_{*}  = \| \m X - \m P_{\v s}\m X\|^2_{F}.
\]
where $\|\m A\|_{*} = \sum _{i=1}^{\min\{m,n\}}\sigma _{i}(\m A)$ for $\m A \in \bb{R}^{m \times n}$ is the trace matrix norm.
This connection is used in \citep{derezinski2020improved} to provide shared approximation bounds for both the CSSP and Nyström approximation. Given that the kernel matrix is always positive semi-definite, the decomposition $\m K = \m X^{\top}\m X$ always exists and one can solve the CSSP for $\m X$ to obtain the best $k$-landmark Nyström approximation with respect to the trace norm. We note that such a decomposition is not unique, e.g., it can be the Cholesky decomposition or the symmetric square-root decomposition.

The matrix $\m X$ does not need explicit evaluation in order to perform CSSP as one can attain $\nabla f_{\lambda}(\v t)$ with the matrix $\m K$ instead (see, \cref{lemma: f_grad}). Therefore, finding the decomposition $\m K = \m X^{\top}\m X$ is not required, and one can approximately solve the CSSP by minimizing $\nabla f_{\lambda}(\v t)$ with the kernel matrix $\m K$.

Suppose instead we want to use the Frobenius matrix norm to find the best choice of columns of the matrix $\m K$ to construct the Nyström approximation. This problem is formulated as
\begin{equation}
\label{eq: nys_exact}
     \argmin_{\v s \in \{0, 1\}^n} \|\m K - \widehat{\m K}_{\v s}\|^2_{F},\quad\textrm{subject to }\|\v s\|_0 \leq k.
\end{equation}

Similar to  $\widetilde{\m P}(\v t)$  we can weight each column $j$ of $\m K$ by $t_j\in[0,1]$ instead of sampling the columns $\m K_{[\v s]}$ for the Nyström approximation.  We define continuous generalization for the Nyström approximation,
\begin{definition}
For $\v t = (t_1,\dots,t_n)^{\top}\in [0,1]^n$ let $\m T := \operatorname{Diag}(\v t)$ and 
\begin{equation*}
\label{Kt}
    \widetilde{\m K}(\v t) :=\m K \m T \left[\m T \m K \m T + \delta(\m I-\m T^2)\right]^{\dagger}\m T\m K,
\end{equation*}
\end{definition}

where $\delta>0$ is a fixed constant. Similar to $\widetilde{\m P}(\v t)$, for any $\v t \in [0,1)^n$ the matrix $\m T \m K \m T + \delta(\m I-\m T^2)$ is invertible. 

In the following two results, we state that $\widetilde{\m K}(\v t)$ is a continuous function on $[0,1]^n$ and agrees with the exact Nyström approximation at every corner point.

\begin{lemma}
\label{lem: cnr_k}
For any corner point $\v s\in\{0,1\}^n$, $\widetilde{\m K}(\v s)$ exists and
\[\widetilde{\m K}(\v s) =  \widehat{\m K}_s
=\m K_{[\v s]}\m K_{[\v s,\v s]}^{\dagger}\m K_{[\v s]}^{\top}.\]
\end{lemma}

\begin{lemma}
\label{lem: cont_k}
$\widetilde{\m K}(\v t)$ is continuous element-wise over $[0,1]^n$. Moreover, for any sequence $\v t^{(1)},\v t^{(2)}\dots\in[0,1)^n$ converging to $\v t\in[0,1]^n$, the limit $\lim_{l\to\infty} \widetilde{\m K}(\v t^{(l)})$ exists and
\[\lim_{l\to\infty} \widetilde{\m K}(\v t^{(l)}) = \widetilde{\m K}({\v t}).\]
\end{lemma}

We therefore have the continuous generalization of the exact problem \eqref{eq: nys_exact},
\begin{equation*}
    \argmin_{\v t \in [0, 1]^n} \|\m K - \widetilde{\m K}(\v t)\|^2_{F},\quad\text{subject to }\sum_{j=1}^n t_j \leq  k.
\end{equation*}
Instead of solving this constrained problem, for a tunable parameter $\lambda$, we consider minimizing the Lagrangian function,
\begin{equation*}
\label{eq: combss_nys}
\argmin_{\v t \in [0, 1]^n} g_{\lambda}(\v t),\quad g_{\lambda}(\v t):= \|\m K - \widetilde{\m K}(\v t)\|^2_{F}+  \lambda\sum_{j=1}^n t_j.
\end{equation*}
As with the continuous extension for CSSP we use a gradient descent method to solve the above problem. The following result provides an expression for $\nabla g_{\lambda}(\v t)$ for $\v t \in (0,1)^n$.
\begin{lemma}
\label{lem: g_grad}
Let $\m Z =\m K - \delta \m I$, $\m L_{\v t} = \m T \m Z \m T +\delta\m I$ and $\m D = \widetilde{\m K}(\v t)-\m K$. Then, for $\v t \in (0,1)^n$,
\begin{equation*}
   \nabla g_{\lambda}(\v t)  = 4\operatorname{Diag}\left[\m L_{\v t}^{-1}\m T \m K \m D\m K \left( \m I  -\m T\m L_{\v t}^{-1}\m T\m Z \right)\right]+\lambda\m 1.
\end{equation*}
\end{lemma}
Evaluating $\nabla g_{\lambda}(\v t)$ has a computational complexity of $O(n^3)$ due to the required inversion of $\m L$ and evaluation of $\m K(\v t)$. As with $\nabla f_{\lambda}(\v t)$ we detail an unbiased estimate for $\nabla g_{\lambda}(\v t)$ in \cref{sec: imp} which utilizes matrix-vector multiplications with $\m K$ and that helps in reducing the computational cost. 
\section{Implementation}
\label{sec: imp}
In this section, we detail how to efficiently solve the continuous problems posed in \cref{sec: cont ext}. In particular, we detail a non-linear transformation that was also used in \citep{moka2022combss} to make both the CSSP and Nyström approximation optimization problems unconstrained. We then show how one can estimate the gradients using MVMs with $\m X$ and $\m K$.
\subsection{Handling Box Constraints \citep{moka2022combss}}
\label{sec: unc-transformation}
The continuous extension of the CSSP and Nyström approximation requires minimizing the functions $f_{\lambda}(\v t)$ and $g_{\lambda}(\v t)$ over $\v t \in [0,1]^n$. We now consider a non-linear transformation to make both optimization problems unconstrained. Consider the mapping $\v t = \v t(\v w)$ given by,
\begin{equation*}
    t_j(w_j) = 1-\exp(-w_j^2),\quad j=1,\dots,n,
\end{equation*}
then if we consider the optimization of continuous CSSP,
\begin{equation*}
   \v w^* =  \argmin_{\v w \in \bb{R}^p} f_{\lambda}(\v t (\v w)),
\end{equation*}
we attain the solution to \eqref{eq: combss} by evaluating $\v t(\v w^*)$. This is true because for any $a,b\in\bb{R}$,
\[
1-\exp(-a^2)<1-\exp(-b^2)\quad\text{if and only if }a^2<b^2.
\]
In vector form the transformation is $\v t(\v w) = \m 1-\exp(-\v w \odot \v w)$ (here $\odot$ denotes element-wise multiplication) and using the chain rule we obtain for $\v w \in \bb{R}^{p}$,
\begin{equation*}
   \frac{\partial f_{\lambda}(\v t (\v w))}{\partial\v w} = \frac{\partial f_{\lambda}(\v t (\v w))}{\partial \v t} \odot (2\v w \odot \exp(-\v w\odot\v w)).
\end{equation*}
We can now implement a gradient descent algorithm to approximately obtain $\v t(\v w^*)$. Using this approximation we can select an appropriate binary vector as a solution to the exact problem \eqref{bcss}. The same transformation can be applied to solve $g_{\lambda}(\v t(\v w))$ over $\v w \in \bb{R}^n$.

\subsection{Stochastic Estimate for the Gradient}
As discussed in \cref{sec: cont ext},  $\nabla f_{\lambda}(\v t )$ and $\nabla g_{\lambda}(\v t )$ are problematic to compute for large $n$ due to the $O(n^3)$ complexity of inverting a matrix. Here we show that we can implement a stochastic gradient descent (SGD) which has strong theoretical guarantees \citep{robbins1951stochastic} by using an unbiased estimate for $\nabla f_{\lambda}(\v t )$ and $\nabla g_{\lambda}(\v t )$.

The method we employ is a factorized estimator $\hat{\ell}$ for the diagonal of a square matrix. Suppose we wish to estimate the diagonal of the matrix $\m A = \m B \m C^{\top}$ where $\m A, \m B, \m C \in \bb{R}^{n\times n}$. Let $\v z \in \bb{R}^n$ be a random vector sampled from the Rademacher distribution, whose entries are either $-1$ or $1$, each with probability $1/2$. Then an unbiased estimate for $\operatorname{Diag}\left(\m A\right)$ is $\hat{\ell}= \m B\v z\odot\m C\v z$, see \citep{martens2012estimating}. Further analysis of its properties including its variance can be found in \citep{mathur2021variance}. We note that when $\m B = \m A$ and $\m C = \m I$, this estimator reduces to the well-known \citep{bekas2007estimator} estimator for the diagonal.

The two following results provide an unbiased estimate for $\nabla f_{\lambda}(\v t )$ and $\nabla g_{\lambda}(\v t )$ using the factorized estimator for the diagonal of a matrix.

\begin{lemma}
\label{lemma: stoch_f}
Recall that in the continuous CSSP  optimization for $\m X$, we have the definitions $\m T = \operatorname{Diag}(\v t)$ for $t\in[0,1]^n$, $ \m K = \m X^{\top}\m X$, $\m Z =\m K - \delta \m I_n$ and $\m L_{\v t} = \m T \m Z \m T +\delta\m I_n$.

Suppose $\v z\in\bb{R}^n$ follows a Rademacher distribution and let:

$(1)\,\v a= \m K\v z,\,(2)\, \v b = \m L_{\v t}^{-1}( \v t\odot \v a)$ and 
\[
\v \phi = \v b \odot \m Z(\v t \odot \v b) - \v a \odot \v b.
\]
Then for $\v t \in (0,1)^n$,
\[\nabla f_{\lambda}(\v t) = 2\bb{E}\left[\v \phi\right]+\lambda \m 1.\]
\end{lemma}
\begin{lemma}
\label{lemma: stoch_g}
Recall that in the continuous Nyström optimization for a kernel matrix $\m K$, we have the definitions  $\m T = \operatorname{Diag}(\v t)$ for $t\in[0,1]^n$, $\m Z =\m K - \delta \m I$ and $\m L_{\v t} = \m T \m Z \m T +\delta\m I$.

Suppose $\v z\in\bb{R}^n$ follows a Rademacher distribution and let:

$(1)\,\v a= \m K \v z,\,(2)\,\v b = \m L_{\v t}^{-1}( \v t \odot \v a),\,(3)\,\v c=\m K(\v t\odot \v b)-\v a,\, (4)\,\v d = \m K \v c,\,(5)\,\v e = \m L_{\v t}^{-1}(\v t \odot \v d)$ and 
\[\v \psi = \v b \odot \v d+\v a \odot \v e - \v e\odot \m Z(\v t \odot\v b)-\v b\odot \m Z(\v t \odot\v e).\]
Then for $\v t \in (0,1)^n$,
\[\nabla g_{\lambda}(\v t) = 2\mathbb{E}\left[\v \psi\right]+\lambda \m 1.\]
\end{lemma}
Using these results, we can obtain for a Monte-Carlo size~$M$, the approximations $\nabla f_{\lambda}(\v t)\approx2\left(\frac{1}{M}\sum_{i=1}^M\v\phi^{(i)}\right)+\lambda\m 1$ and $\nabla g_{\lambda}(\v t)\approx2\left(\frac{1}{M}\sum_{i=1}^M\v\psi^{(i)}\right)+\lambda\m 1$, where $\v \phi^{(i)}$ and $\v \psi^{(i)}$ are evaluated using a sample $\v z^{(i)}$ drawn from the Rademacher distribution.

These results show that to evaluate stochastic gradients one needs to solve linear systems efficiently with the matrix $\m L_{\v t}$. These systems can be iteratively solved using the conjugate gradient (CG) algorithm \citep{golub1996matrix} which uses a sequence of MVMs with $\m L_{\v t}$. Multiplying a vector with $\m L_{\v t}$ can be reduced to a single MVM with the matrix $\m K$ and a sequence of element-wise vector multiplications and additions.

\subsection{Obtaining a Solution}
\label{sec: obt_sol}

\begin{figure*}[ht]
\centering
\begin{subfigure}{.45\textwidth}
  \centering
  \includegraphics[width=\linewidth]{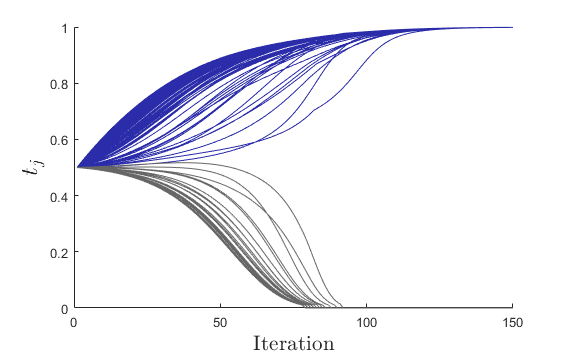}  
  \caption{Gradient Descent}
\end{subfigure}
\begin{subfigure}{.45\textwidth}
  \centering
  \includegraphics[width=\linewidth]{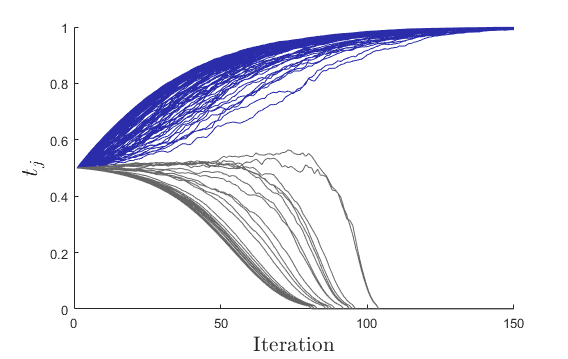}  
  \caption{Stochastic Gradient Descent}
\end{subfigure}
\caption{Convergence of $\v t$ for continuous Column Subset Selection using the MNIST dataset. Blue trajectories correspond to selected columns. Only a subset of 300 randomly chosen column trajectories (out of 784) are displayed. For both (a) and (b), $\lambda = 10$ and $\delta = 10$. In (b) the Monte-Carlo
size is $M=5$.}
\label{fig: sol_path}
\end{figure*}

While we have re-framed both the CSSP and the Nyström problem as an optimization over $\v t \in [0,1]^n$, the priority remains to obtain an approximate solution $\v s \in \{0,1\}^n$ to \eqref{bcss} and \eqref{eq: nys_exact}. To obtain such a binary vector, we first initialize SGD from a starting point $\v t^{(0)}$ and return the final value $\v t^*$ after a termination condition for SGD has been satisfied. Under SGD the iterative sequence $\{\v t^{(i)}\}_{i\geq 0}$ moves towards a corner point of the hypercube. To obtain the closest corner point $\v s \in \{0,1\}^n$, we map the insignificant $t^*_j$'s to $0$ and  all the other $t^*_j$'s to 1 for some tolerance parameter $\tau\in(0,1)$. This implementation is shown \cref{alg: cont_landm}. In \cref{fig: sol_path} we provide example solution paths $\{\v t^{(i)}\}_{i\geq 0}$ under both batch gradient descent and SGD.

When choosing the value for $\v t^{(0)}$ it is important to consider the following true statements: $t_j = 0$ if and only if $w_j = 0$ and 
\[\lim_{w_j\to 0}\frac{\partial f_{\lambda}(\v t (\v w))}{\partial w_j}=\lim_{w_j\to 0}\frac{\partial g_{\lambda}(\v t (\v w))}{\partial w_j}=0.\]
These facts imply that if $t_j$ is set to zero during the course of the optimization it will remain unchanged thereafter. Therefore, it is important to choose $\v t^{(0)}$ that is away from any corner point. It is for this reason,  we set $\v t^{(0)} = (1/2,\dots , 1/2)^{\top}$ in all our experiments.

\begin{algorithm}[tb]
   \caption{Continous Landmark Selection}
   \label{alg: cont_landm}
\begin{algorithmic}[1]
   \STATE {\bfseries input: }Data matrix: $\m X \in \bb{R}^{m\times n}$ (CSSP) or Kernel matrix: $\m K \in \bb{R}^{n\times n}$ (Nyström method), Tuning parameters: $\delta$ and $\lambda$, Monte Carlo size: $M$, Termination Condition: $\operatorname{TermCond}$, Threshold value: $\tau\in[0,1]$.
   \STATE Set  $\v t^{(0)} = (1/2,\dots\, 1/2)^{\top}$
   \STATE  $\v w^{(0)}\leftarrow \sqrt{-\ln(1-\v t^{(0)})}$
   \STATE  $\v w^*\leftarrow$ SGD $(\v w^{(0)}$, $M$, $\m X$  or $\m K$, $\operatorname{TermCond}$)
   \STATE $\v t^*\leftarrow1-\exp(-\v w^*\odot\v w^*)$
   \FOR{$i=1$ {\bfseries to} $n$}
   \STATE $s_j\leftarrow  I(t^*_j>\tau)$\;
   \ENDFOR
   \STATE{\bfseries return: }$\v s^* = (s_1,\dots,s_n)^{\top}$
\end{algorithmic}
\end{algorithm}

\subsection{Dimensionality Reduction}
\label{sec: dim_red}
In \cref{sec: obt_sol} we stated that if $t_j$ is set to zero during the course of the SGD then it will remain unchanged thereafter. This opens the possibility to reduce the computational cost of estimating $\nabla f_{\lambda}(\v t(\v w))$ and $\nabla g_{\lambda}(\v t(\v w))$ by only focusing on terms where $t_j \neq 0$.

Let $\c N = \{1,\dots,n\}$ and for any $\v t \in [0,1)^n$ let $\c I_{\v t}= \{j:  t_j = 0\}$. For a vector $\v a\in \bb{R}^n$, denote the vector $(\v a)_+$ of dimension $n-|\c I_{\v t}|$ (respectively $|\c I_{\v t}|$) constructed from $\v a$ by removing the elements with indices that are in $\c I_{\v t}$ (respectively, in $\c N  \setminus \c I_{\v t}$). Likewise, for a matrix $\m A\in \bb{R}^{n\times n}$, denote the principal sub-matrix $(\m A)_+$ (respectively, $(\m A)_0$) that is constructed by removing the rows and columns with indices that are in $|\c I_{\v t}|$ (respectively, in $\c N  \setminus \c I_{\v t}$). Then, we have the following result.

\begin{lemma}
\label{lem: dim_red}
For any expression $\v q = \m L^{-1}_{\v t}(\v t \odot \v r)$ where $\v r \in \bb{R}^{n}$ and $\v t \in [0,1)^n$,
\[
(\v q)_0 = \v 0 \quad\text{and}\quad (\v q)_{+} = \left((\m L_{\v t})_+\right)^{-1}\left((\v t)_{+}\odot (\v r)_{+}\right),
\]
where,
\[
(\m L_{\v t})_+ = (\m T)_{+}(\m K)_{+}(\m T)_{+}+\delta(\m I - (\m T)^2_{+}).
\]
\end{lemma}
To incorporate this result in our algorithm, we set a small constant $\epsilon$ and during the course of SGD if $\v t (\v w)_j<\epsilon$, we set its value to zero. Thereafter, when solving $(2)$ and $(5)$ in either \cref{lemma: stoch_f} (CSSP) or \cref{lemma: stoch_g} (Nyström) the dimension of the linear system is $n-|\c I_{\v t}|< n$.

\begin{figure*}[ht]
\centering
\begin{subfigure}{.24\textwidth}
  \centering
  \includegraphics[width=\linewidth]{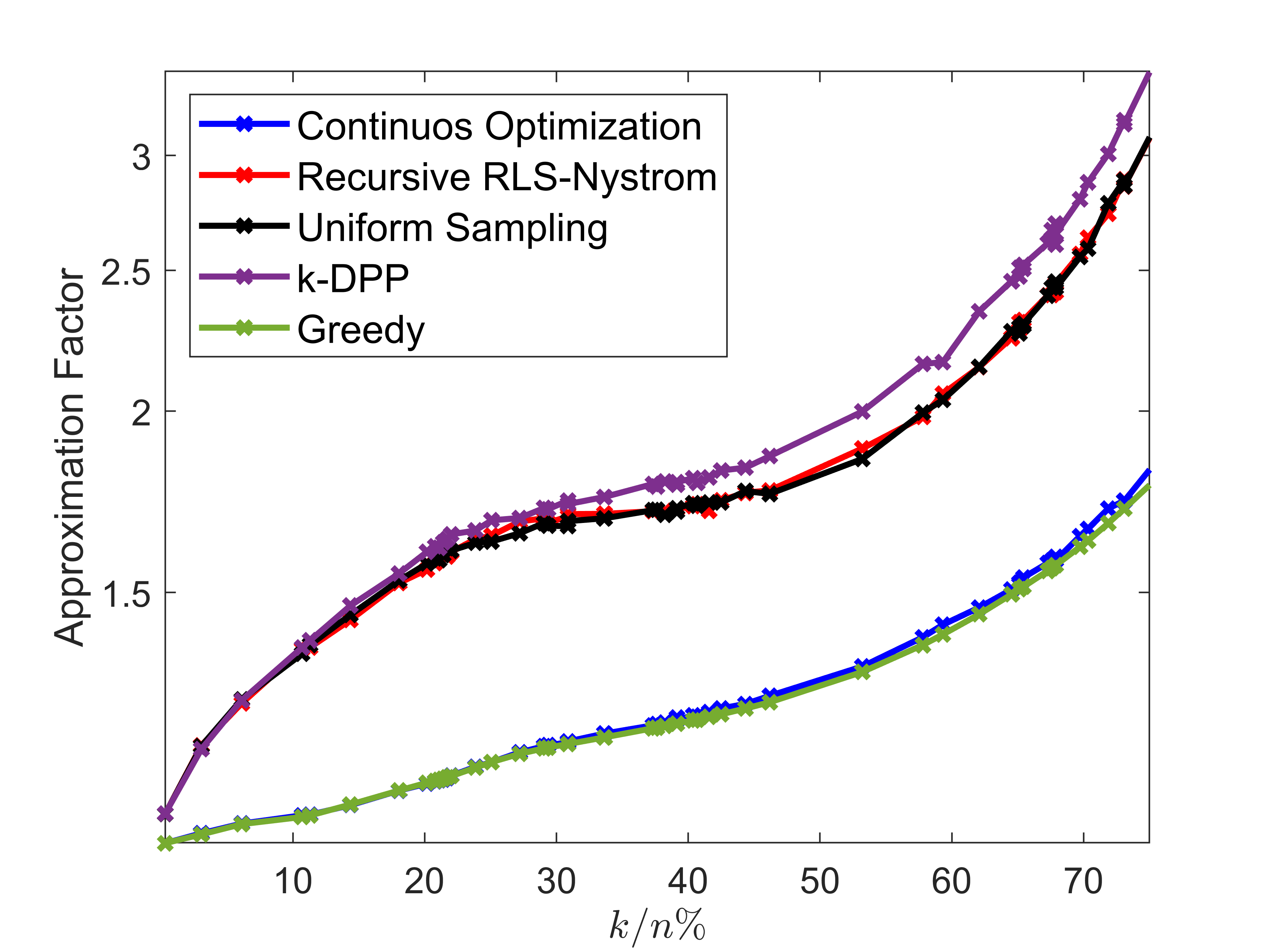}   
  \caption{Residential ($\sigma = 1$)}
\end{subfigure}
\begin{subfigure}{.24\textwidth}
  \centering
  \includegraphics[width=\linewidth]{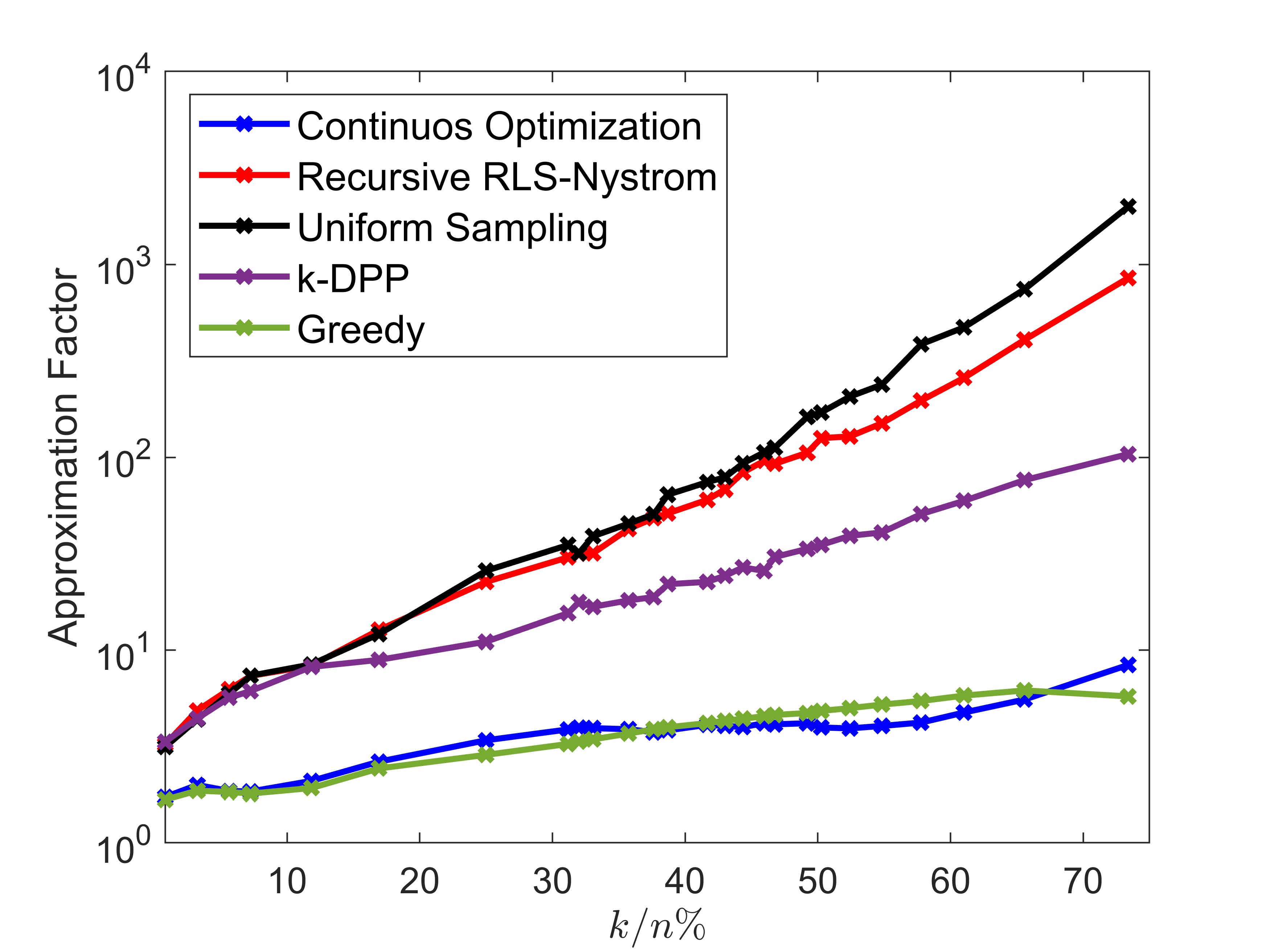}   
  \caption{Residential ($\sigma = 5$)}
\end{subfigure}
\begin{subfigure}{.24\textwidth}
  \centering
  \includegraphics[width=\linewidth]{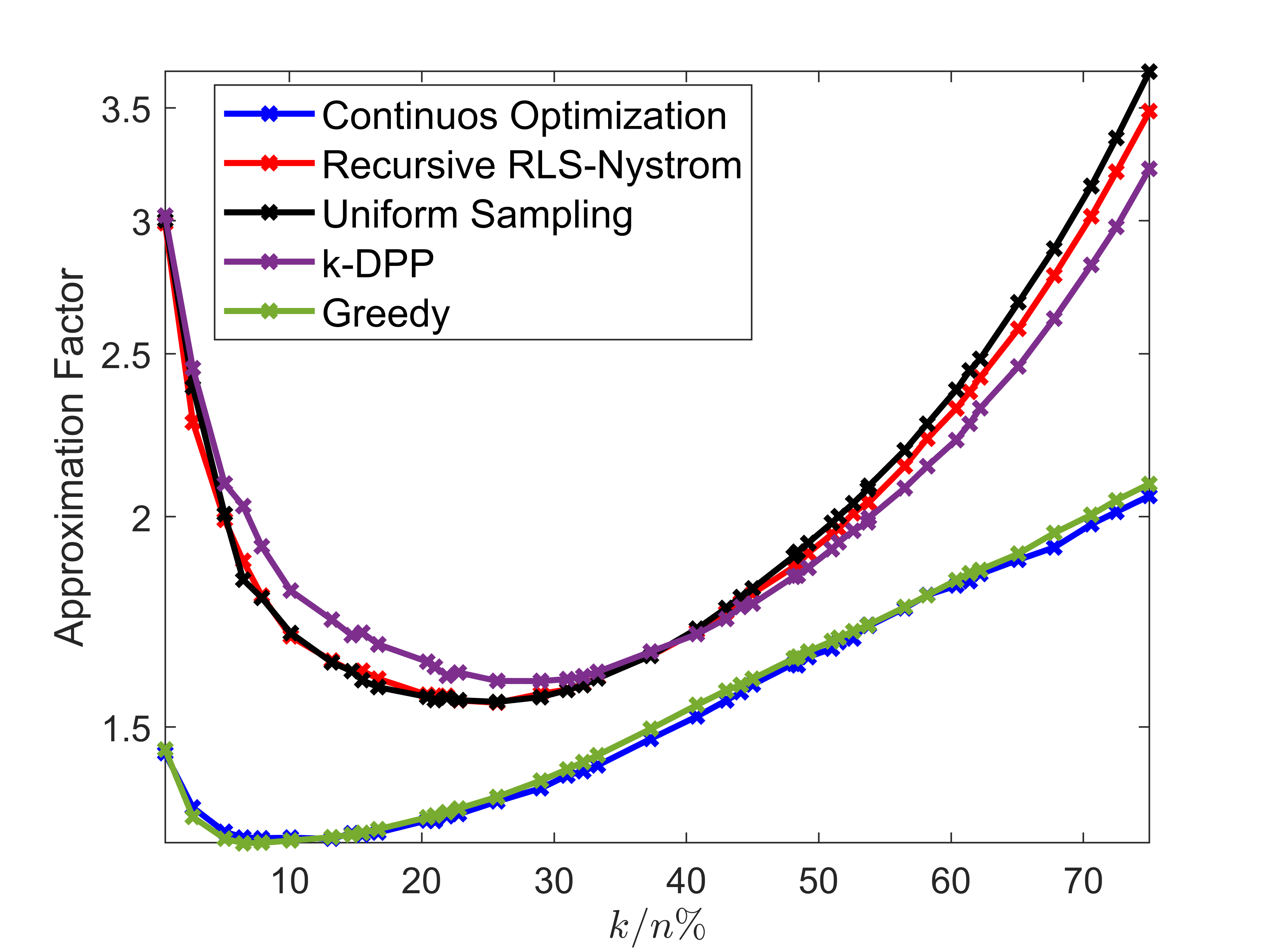}  
  \caption{MNIST1K ($\sigma = 10$) }
  \label{fig:sub-mnist10}
\end{subfigure}
\begin{subfigure}{.24\textwidth}
  \centering
  \includegraphics[width=\linewidth]{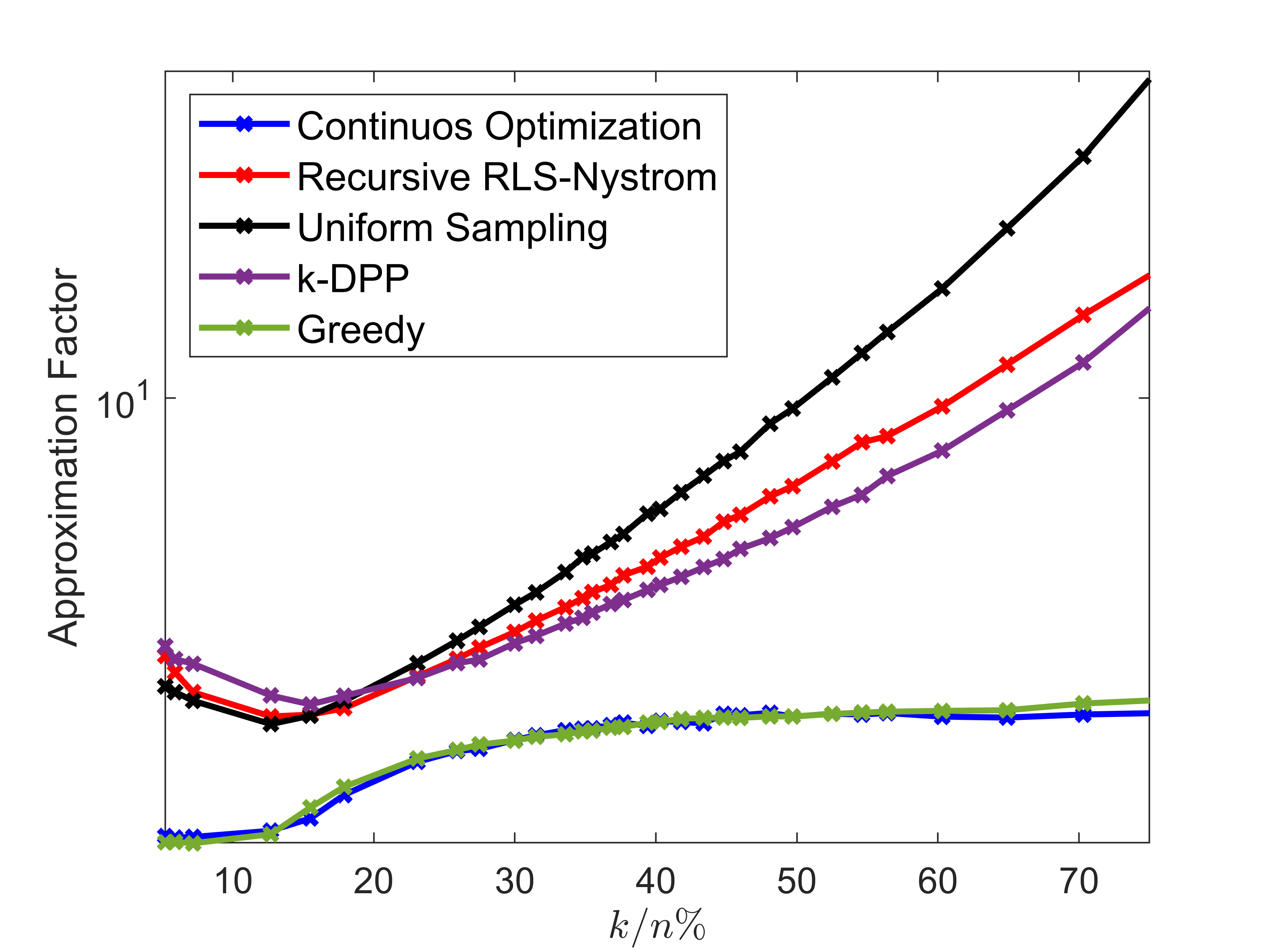}  
  \caption{MNIST1K ($\sigma = 20$)}
  \label{fig:sub-mnist20}
\end{subfigure}
\caption{The mean Nyström empirical approximation factor over 50 trials for the UCI Residential Building and MNIST dataset where $\m K$ is constructed using the Gaussian Radial Basis Function (RBF) kernel:  $\m K_{i,j} = h(\v x'_i, \v x'_j ) = \exp\left(-\|\v x'_i- \v x'_j\|^2\right)/\sigma^2$. Approximation factor is plotted on a logarithmic scale.}
\label{fig: smallscalenys}
\end{figure*}

\begin{figure*}[ht]
\centering
\begin{subfigure}{.24\textwidth}
  \centering
  \includegraphics[width=\linewidth]{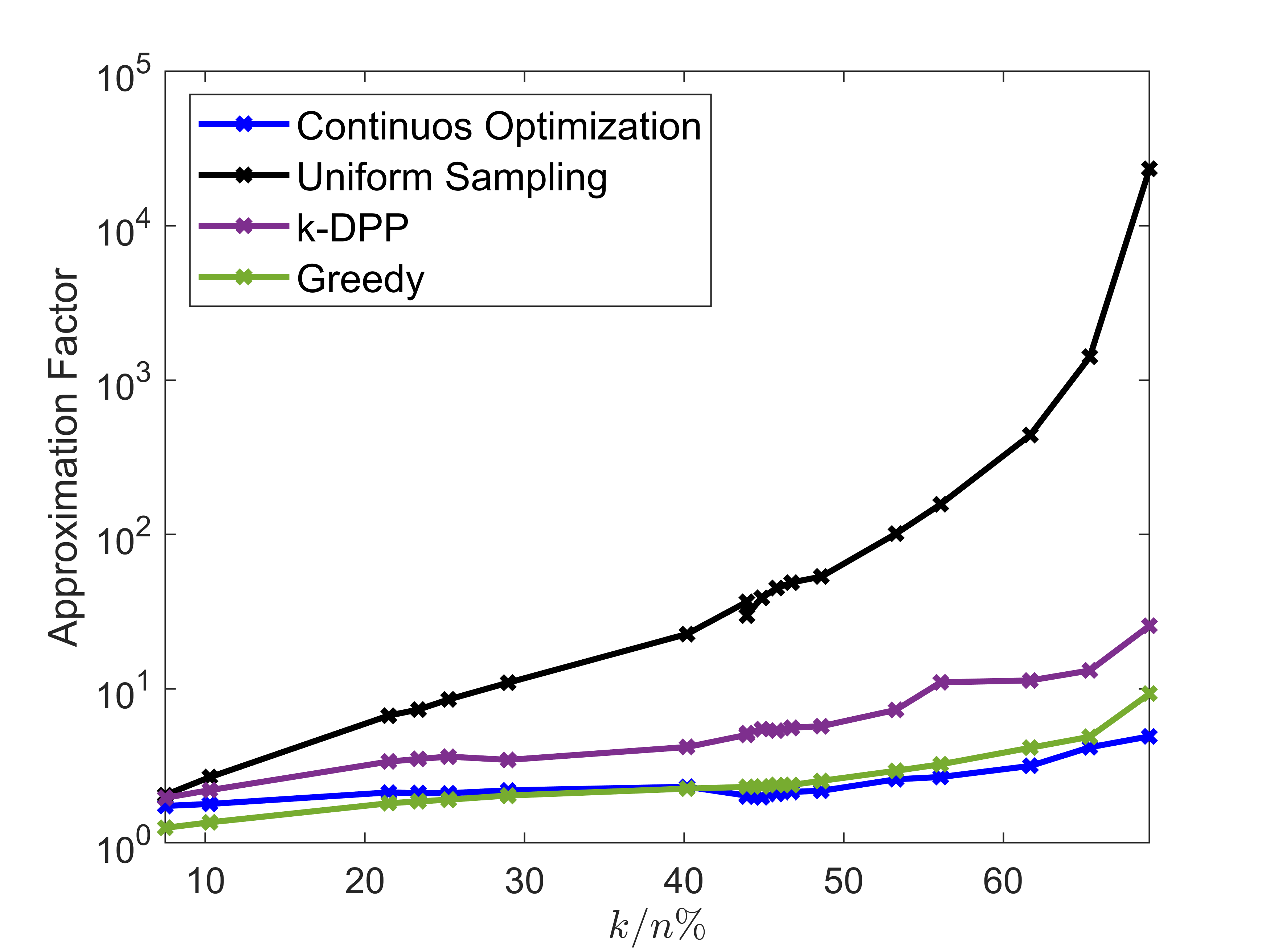}   
  \caption{Residential}
\end{subfigure}
\begin{subfigure}{.24\textwidth}
  \centering
  \includegraphics[width=\linewidth]{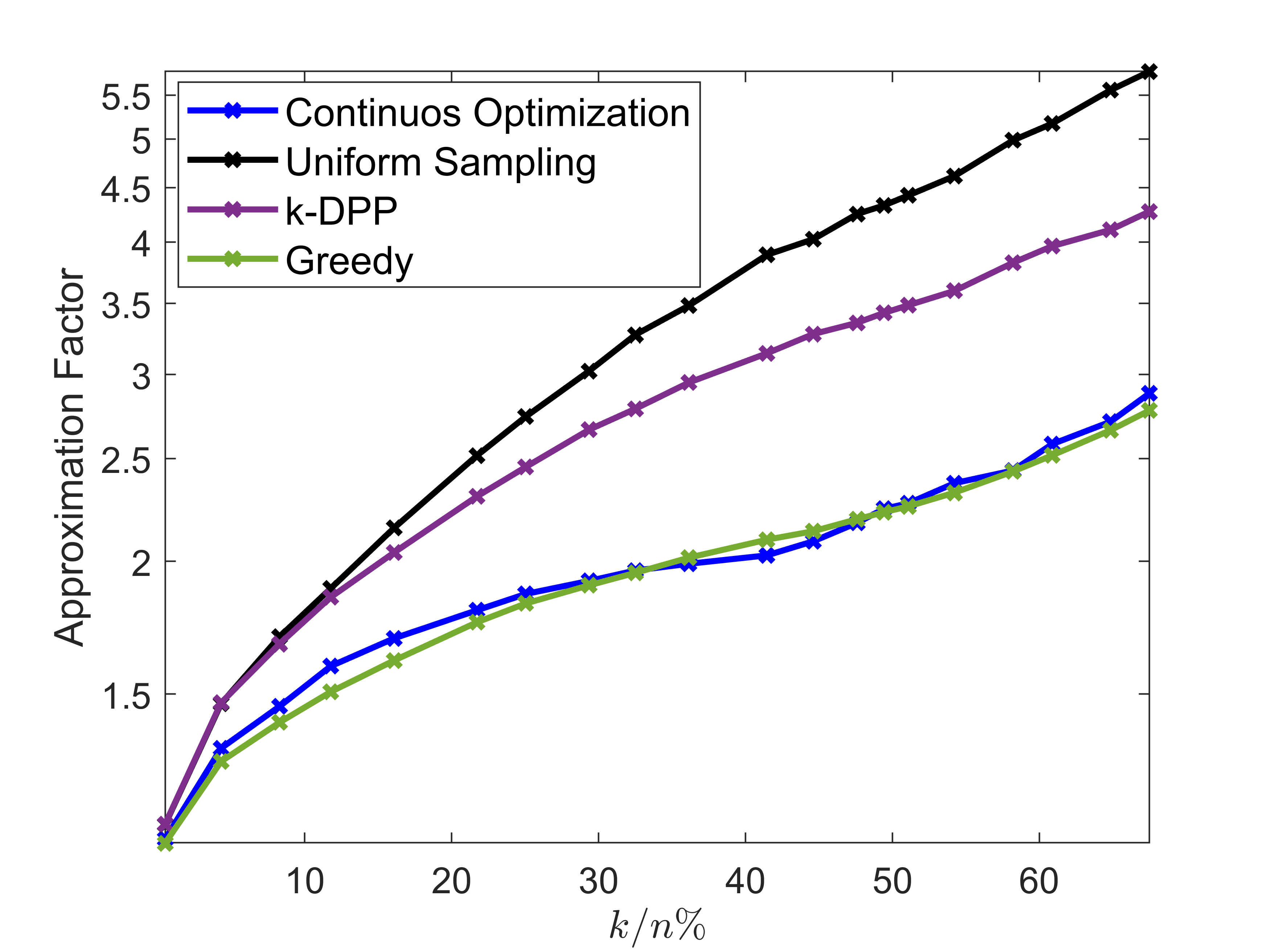}   
  \caption{MNIST1K}
\end{subfigure}
\begin{subfigure}{.24\textwidth}
  \centering
  \includegraphics[width=\linewidth]{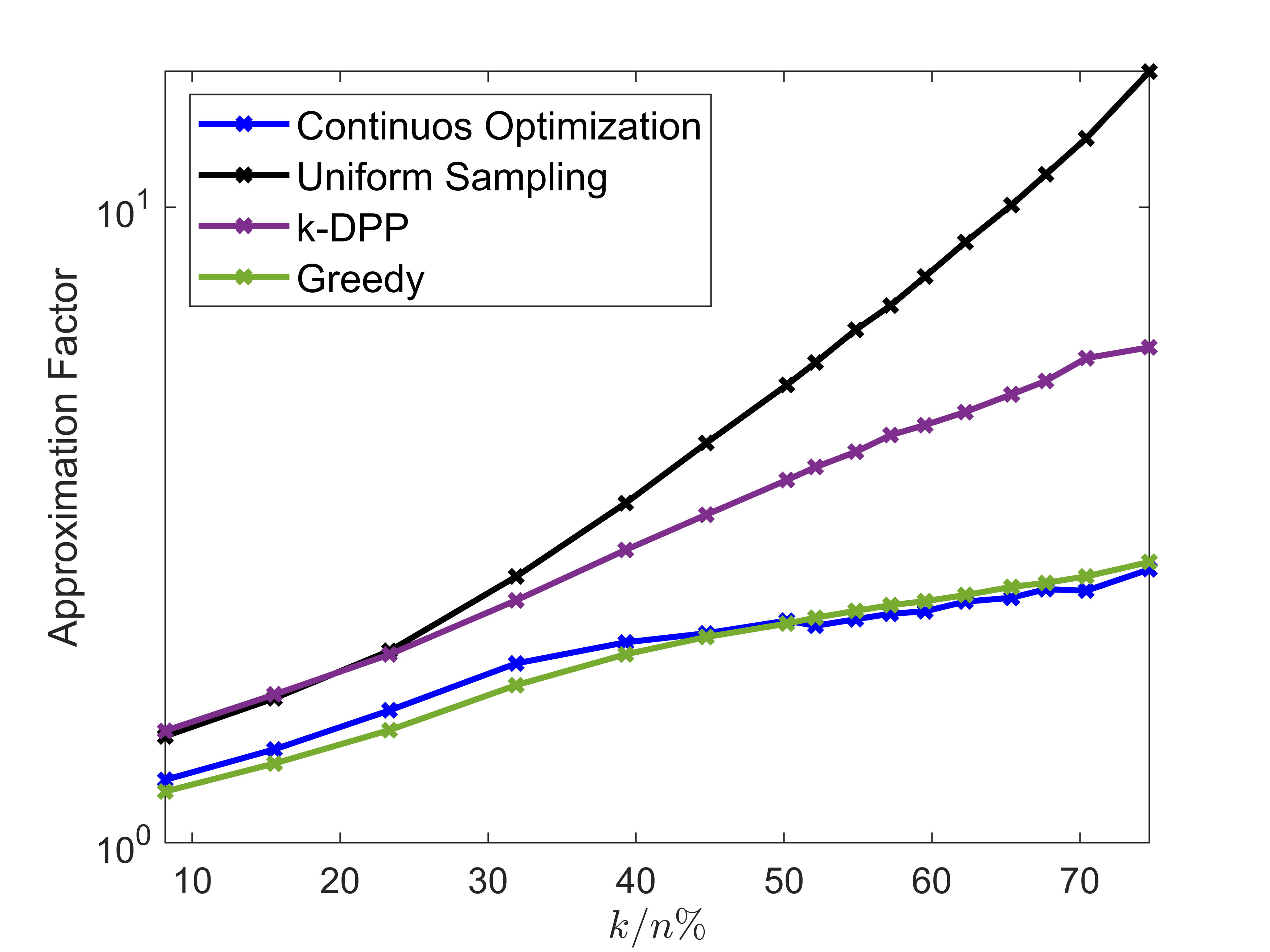}  
  \caption{Arrhythmia}
\end{subfigure}
\begin{subfigure}{.24\textwidth}
  \centering
  \includegraphics[width=\linewidth]{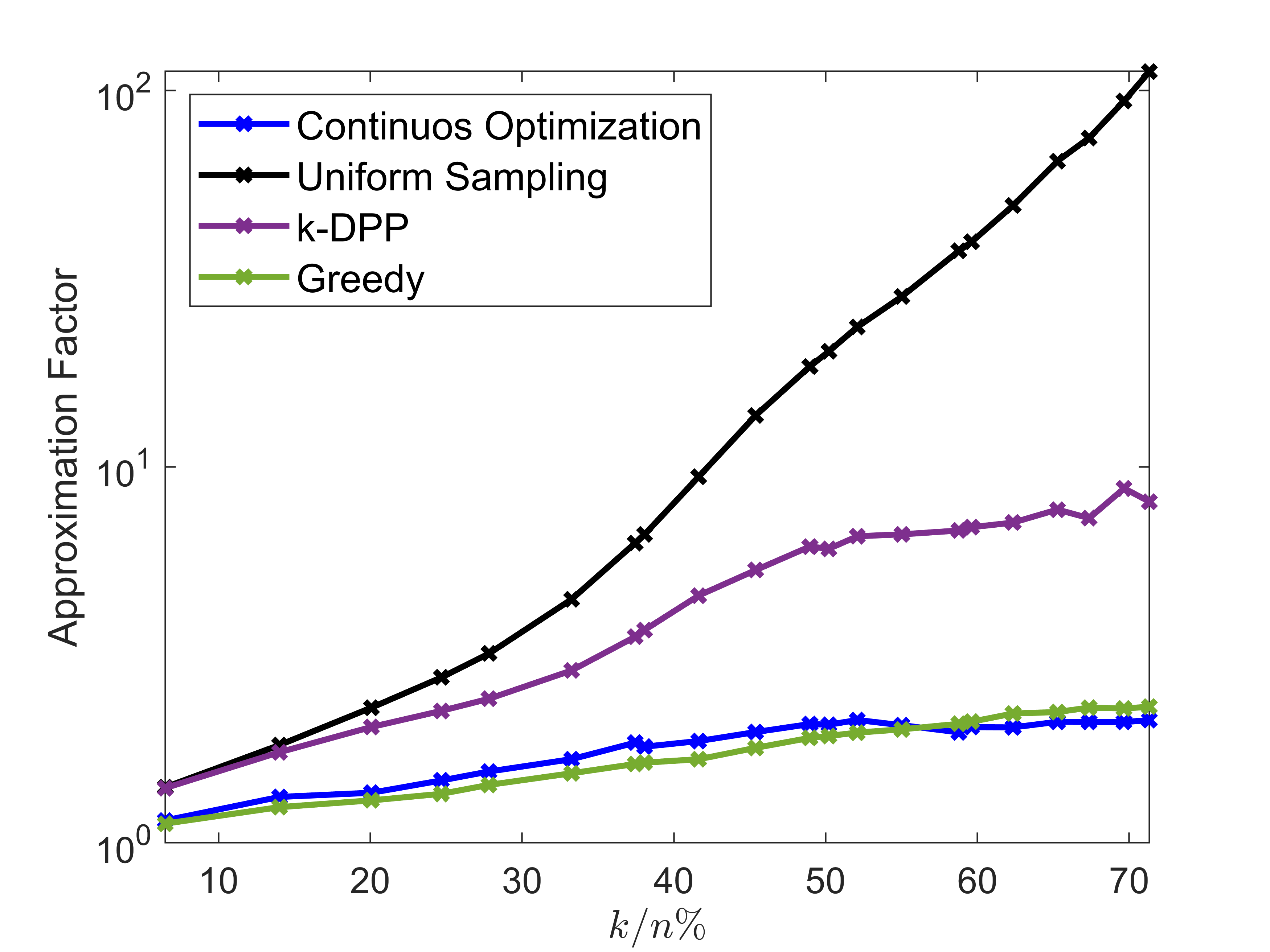}  
  \caption{SECOM}
\end{subfigure}
\caption{The mean CSSP empirical approximation factor over 50 trials for the MNIST dataset and three UCI datasets for different methods. Approximation factor is plotted on a logarithmic scale.}
\label{fig: smallscalecssp}
\end{figure*}
\subsection{Complexity Analysis}
The main computational cost of our algorithm is the complexity attributed to estimating the gradients at each iteration of SGD. For simplicity of analysis, we assume the dimensionality reduction described in \cref{sec: dim_red} is not carried out.  The cost to solve $(2)$ and $(5)$ in either Lemma \ref{lemma: stoch_f} or \ref{lemma: stoch_g} via CG is $O(T_{mult} M\ell)$ flops where $\ell$ is the number of CG iterations and $T_{mult}$ is the cost of computing a matrix-vector product with either $\m X^{\top}\m X$ (CSSP) or kernel matrix $\m K$ (Nyström). Generally, only $\ell \ll n$ iterations of CG are required to obtain an accurate solution to the linear system.

The cost $ T_{mult}$ is $O(mn)$ and $O(n^2)$ via direct computation for $\m X ^{\top}\m X$ and $\m K$ respectively. For kernel matrices with specific structure, this cost can be reduced. For example, for Toeplitz matrices or for matrices constructed from a kernel function that is analytic and isotropic, the cost can be reduced to quasi-linear complexity \citep{dietrich1997fast, gardner2018product, ryan2022fast}.  Utilizing GPU hardware for accelerating matrix computations has gained significant recent attention and numerous software regimes \citep{charlier2021kernel, hu2022giga} have been proposed to accelerate kernel MVMs. These methods can be implemented out-of-the-box and allow MVMs to be feasible on very large datasets ($n\sim10^8$). Another advantage of these algorithms is that, as long as the kernel function $h(\v x'_i, \v x'_j )$ is given, MVMs can be computed directly without ever storing the kernel matrix $\m K$. This is an advantage of our method when compared to other methods such as the greedy selection method for the Nyström approximation in \citep{farahat2011novel}, which has a cost of $O(n^2k)$ and requires the full explicit matrix to be stored in memory.

\subsection{Role of parameters $\delta$ and $\lambda$}

The tuning parameter $\lambda$ controls the size of the penalty $\|\v t\|_1$ which is added to the Frobenius matrix loss. It is intuitive then that for a larger value of $\lambda$ a stronger shrinkage is applied to $\v t$ during the course of the continuous optimization. In terms of curvature, as $\lambda$ increases so does the directional slope of $f_{\lambda}(\v t (\v w))$ and $g_{\lambda}(\v t (\v w))$ in the region around $w_j = 0$. For this reason, it is likelier that more $w_j$'s will be pushed towards zero when the value for $\lambda$ is large. This behavior is similar to that of the parameter $\lambda$ in the COMBSS method \citep{moka2022combss} where a more formal analysis can be found.  We note that the relationship between $\lambda$ and $k$ is data dependent and it is suggested that the user apply an efficient grid search regime to obtain an appropriate $\lambda$ for their use.

With respect to the parameter $\delta$ we first note that \cref{lem: cnr_p} and \cref{lem: cnr_k} remain true regardless of the choice of $\delta$. Therefore, the value of $\delta$ affects the behavior of the penalized loss only at the interior points $\v t\in(0,1)^n$. We would like a choice of $\delta$ such that for all the interior points $\v t\in(0,1)^n$ the functions $f_{\lambda}(\v t)$ and $g_{\lambda}(\v t)$ are well-behaved. When $\delta$ is very small the linear systems that require solving at $\v t\in(0,1)^n$ may be close to singular and numerical issues can arise more frequently. Moreover, when $\delta$ is large we observe large shifts in the value of the objective approaching a corner point. Our simulations indicate that $\delta =1$ produces a well-behaved function.

\begin{figure*}[ht]
\centering
\begin{subfigure}{.33\textwidth}
  \centering
  \includegraphics[width=\linewidth]{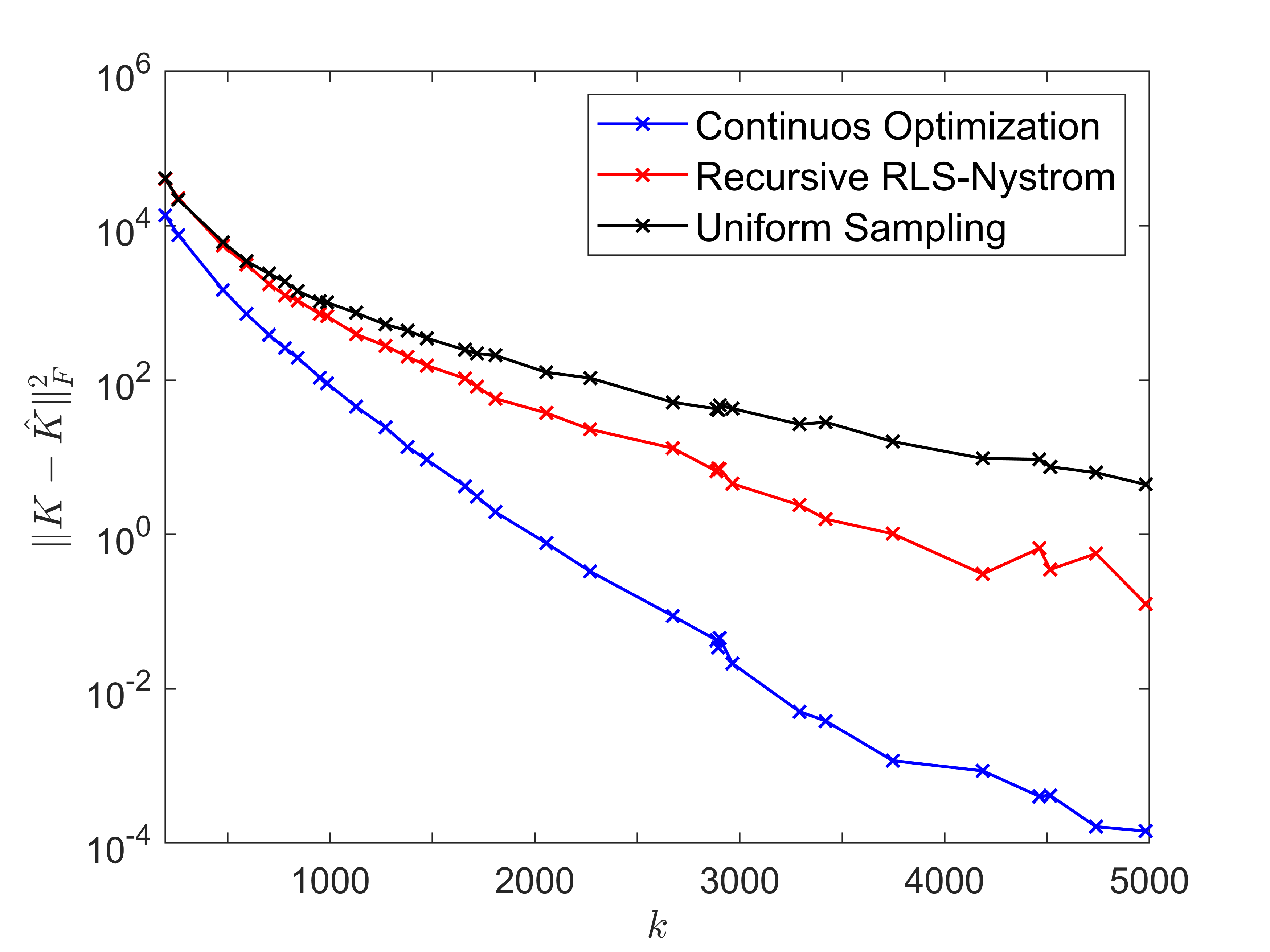}   
  \caption{Power Plant}
\end{subfigure}
\begin{subfigure}{.33\textwidth}
  \centering
  \includegraphics[width=\linewidth]{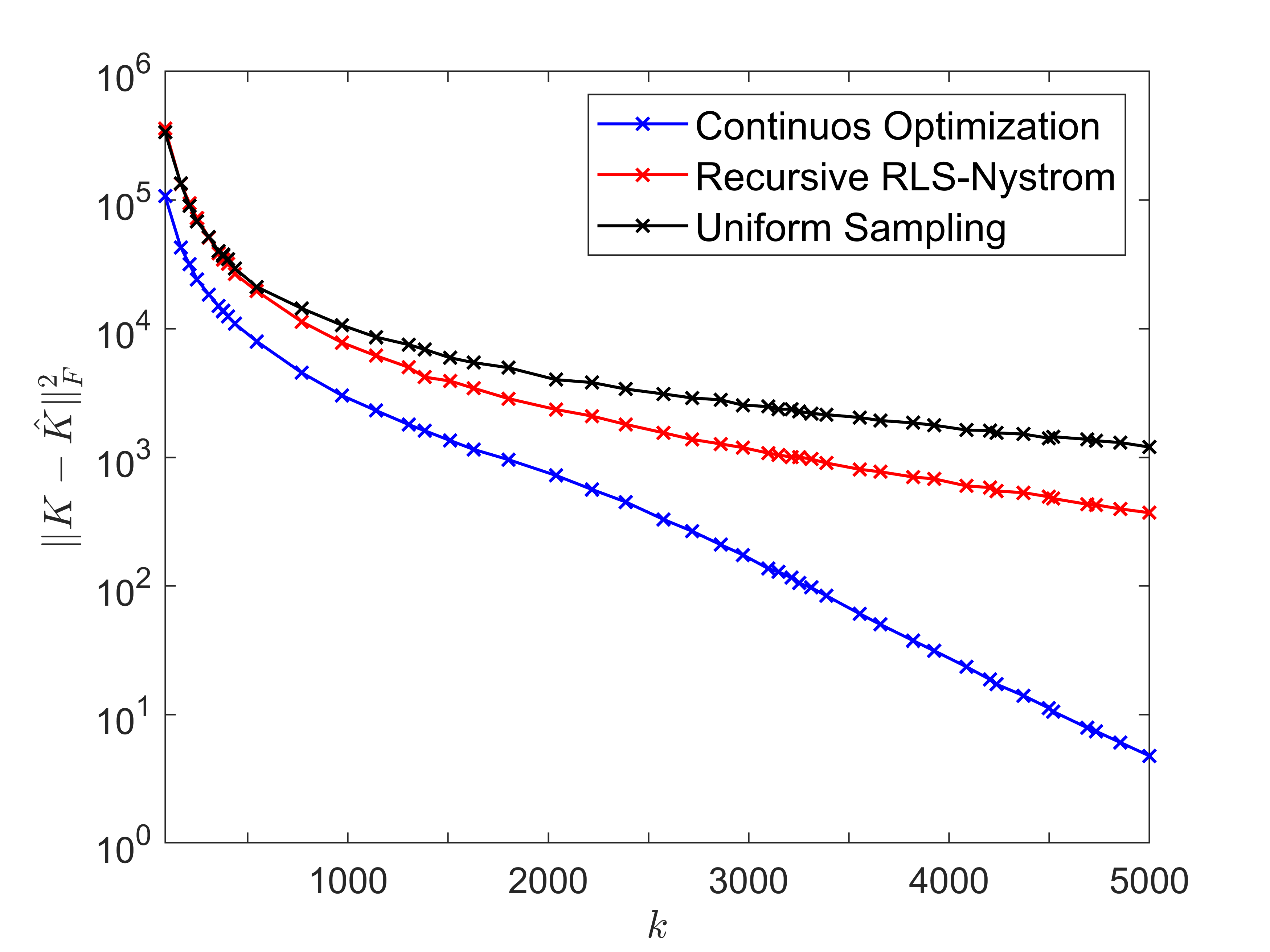}   
  \caption{HTRU2}
\end{subfigure}
\begin{subfigure}{.33\textwidth}
  \centering
  \includegraphics[width=\linewidth]{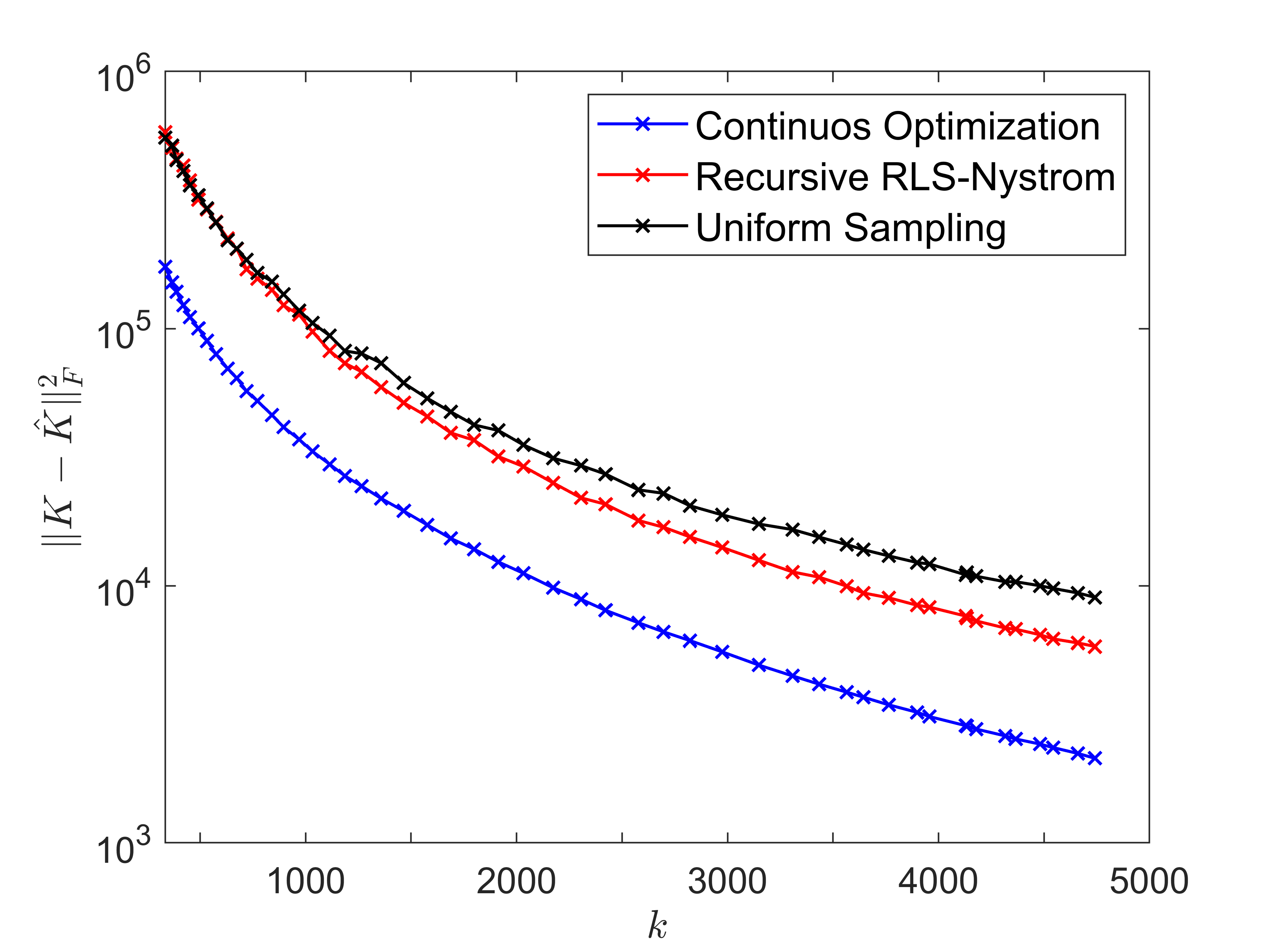}  
  \caption{Protein}
\end{subfigure}
\caption{The mean empirical squared Frobenius error $\|\m K - \widehat{\m K}_{\v s^*}\|^2_{F}$ over 10 trials for the UCI datasets Power Plant, HTRU2 and Protein for different methods. The kernel matrix $\m K$ for all datasets is constructed using the RBF kernel function with $\sigma = 0.5$. Error is plotted on a logarithmic scale.}
\label{fig: largescale}
\end{figure*}

\section{Numerical Experiments and Results}
\label{sec: emp_eval}
In this section, we provide numerical examples with real data designed to demonstrate that our proposed continuous optimization method outperforms well-known sampling-based methods for small and large datasets. Moreover, we demonstrate that when it is feasible to run greedy selection, our continuous method exhibits very similar performance.

Numerical experiments were conducted on the small to medium-sized datasets: Residential and Building dataset ($m = 372$, $n = 109$), MNIST1K ($m = 1000$,  $n = 784$)\footnote{https://yann.lecun.com/exdb/mnist/},  Arrhythmia dataset ($m = 452$, $n = 279$), SECOM ($m = 1567$,  $n = 591$). Numerical experiments for Nyström landmark selection were also conducted on the larger datasets: Power Plant dataset ($m = 4$, $n = 9568$), HTRU2 dataset ($m = 8$, $n = 17898$) and Protein dataset ($m = 9$, $n = 45730$). All datasets except MNIST are downloaded from UCI ML Repository \cite{asuncion2007uci}. All datasets were standardized such that all columns had mean zero and variance equal to one.

For the small to medium-sized datasets, we use the best rank-k approximation factor to compare our method to existing methods (see \cref{fig: smallscalenys} and \cref{fig: smallscalecssp}). The best rank-k approximation factor is given by
\[
\text{Approximation Factor} := \frac{\|\m A -\widehat{\m A}_s\|_F^2}{\|\m A -\widehat{\m G}\|_F^2}.
\]
where $\widehat{\m A}_s$ is either the Nyström or CSSP low-rank matrix and $\widehat{\m G}$ is the best rank-k approximation computed using the Singular Value Decomposition (SVD) of $\m A$.

In these experiments, we compare the proposed continuous landmark selection method executed with SGD ($M = 10$) with the following four well-known methods: Uniform Sampling \citep{williams2000using}, Recursive RLS (Ridge Leverage Scores) - Nyström sampling \citep{musco2017recursive}, k-DPP sampling \citep{derezinski2021determinantal} and Greedy selection \citep{farahat2011novel,farahat2013efficient}.

For the experiments conducted on the larger datasets (see \cref{fig: largescale}) we exclude the k-DPP sampling and greedy methods as it is either too costly to compute the choice of landmark points or too costly to store the full kernel matrix on a GPU. In our implementation of continuous Nyström landmark selection, we use the KeOps library \citep{charlier2021kernel} to efficiently compute MVMs and linear solves on a GPU without ever storing the matrix $\m K$, thus negating the need to store any $O(n^2)$ objects. These experiments were run using an NVIDIA Tesla T4 GPU with 16GB memory.

In \cref{fig: smallscalenys} and \cref{fig: smallscalecssp} we observe the approximation factor for Nyström and CSSP landmark selection with different subset sizes $k$. A lower approximation factor indicates a better approximation and an approximation factor close to one implies near-best-case performance for the given subset size $k$.  The results indicate that the continuous optimization method is better than every tested sampling method and is very similar to greedy selection in performance (whenever the greedy selection is feasible). In most cases, for the CSSP, as the proportion of selected columns increases the continuous method starts to marginally outperform the greedy method.

In \cref{fig: largescale}, we observe for all three datasets (Power Plant, HTRU2 and Protein) that the continuous landmark selection achieves better accuracy than the Recursive RLS (Ridge Leverage Scores) - Nyström sampling and Uniform sampling methods. While Recursive RLS sampling (complexity: $O(nk^2)$) and uniform sampling are faster at selecting landmark points, for a fixed $k$ the continuous method obtains a more accurate Nyström approximation. Thus, if a memory budget for the size of the Nyström approximation is given, as is often the case, the continuous method will compute a superior approximation.

\section{Conclusion}
In this paper, we have introduced a novel algorithm that exploits unconstrained continuous optimization to select columns for both the CSSP and Nyström approximation. The algorithm selects columns by minimizing an extended objective which is defined over the hypercube $[0,1]^n$ rather than iterating over the corner points of the hypercube which correspond to all of the $n\choose k$ subsets. The extended objective for both the CSSP and Nyström approximation can be minimized via SGD where the gradients are estimated with an unbiased estimator which requires only MVMs with either $\m X$ (CSSP) or $\m K$ (Nyström). On the real-world examples that we considered in this article, the proposed method has proven to be more accurate without incurring higher computational cost.

\bibliography{bibliography}
\bibliographystyle{icml2023}

\newpage
\appendix
\onecolumn
\section{Proofs}

\emph{Proof of \cref{lem: cnr_p}}. 
The following proof follows similar arguments to that of \emph{Theorem 1} in \citep{moka2022combss}. Given that the pseudo-inverse of a matrix after a permutation of rows (respectively, columns) is identical to the matrix obtained by applying the same permutation on columns (respectively, rows) on the  pseudo-inverse, we assume without loss of generality that all the zero-elements $\v s \in \{0,1\}^n$ appear at the end, in the form,
\[
\v s = (s_1,\dots\,s_l,0,\dots, 0).
\]
where $l$ is equal to the number of non-zeros in $\v s \in \{0,1\}^n$. Then, $\widetilde{\m P}(\v s)$ is given by the block-wise matrix,

\begin{equation}
\label{ps_1}
\widetilde{\m P}(\v s) = \begin{bmatrix} \m X_{[\v s]}  & \m 0\end{bmatrix}\begin{bmatrix} \m X_{[\v s]}^{\top}\m X_{[\v s]}  & \m 0\\ \m 0 & \delta \m I\end{bmatrix}^{\dagger}\begin{bmatrix} \m X_{[\v s]}^{\top}  \\ \m 0\end{bmatrix}.
\end{equation}
It is easy to verify, when the matrices $\m A_1$ and $\m A_2$ are square,  the block-diagonal pseudo-inverse $\begin{bmatrix} \m A_1  & \m 0\\ \m 0 & \m A_2\end{bmatrix}^{\dagger} = \begin{bmatrix} \m A_1^{\dagger}  & \m 0\\ \m 0 & \m A_2^{\dagger}\end{bmatrix}$. Therefore \eqref{ps_1} reduces to, 
\begin{align*}
\widetilde{\m P}(\v s) &= \begin{bmatrix} \m X_{[\v s]}  & \m 0\end{bmatrix}\begin{bmatrix} \m X_{[\v s]}^{\top}\m X_{[\v s]}^{\dagger}  & \m 0\\ \m 0 & \delta^{-1} \m I\end{bmatrix}\begin{bmatrix} \m X_{[\v s]}^{\top}  \\ \m 0\end{bmatrix}\\
&= \m X_{[\v s]}\left(\m X_{[\v s]}^{\top}\m X_{[\v s]}\right)^{\dagger} \m X_{[\v s]}^{\top} \\
&= \m X_{[\v s]}\m X_{[\v s]}^{\dagger}.
\end{align*}$\square$

\emph{Proof of \cref{lem: cnr_p}}. 
To obtain the gradient $f_{\lambda}(\v t)$ for $\v t \in (0,1)^n$ we first simplify the term $-\tr\left[\m X^{\top}\widetilde{\m P}(\v t)\m X\right]$ by letting $\m K = \m X^{\top}\m X$, $\m Z =\m K - \delta \m I$ and $\m L_{\v t} = \m T \m Z \m T +\v\delta\m I$. Then, we have,
\[
-\tr\left[\m X^{\top}\widetilde{\m P}(\v t)\m X\right] = -\tr\left[\m K\m T\m L^{-1}_{\v t}\m T\m K\right].
\]
Using matrix calculus, for any $j = 1,\dots,n$, we have the partial derivative,
\begin{equation}
    \label{ap:1}
    \frac{\partial}{\partial t_j}\left[-\tr\left(\m X^{\top}\widetilde{\m P}(\v t)\m X\right)\right] = -\tr\left(\m K\frac{\partial\left[\m T\m L^{-1}_{\v t}\m T\right]}{\partial t_j} \m K\right).
\end{equation}
Let $\m E_j$ be the square matrix of dimension $n \times n$ with 1 at position $(j,j)$ and 0 everywhere else. Then $\frac{\partial\m T }{\partial t_j} = \m E_j$ and we have,
\begin{equation}
    \label{ap:2}
    \frac{\partial\left[\m T\m L^{-1}_{\v t}\m T\right]}{\partial t_j} = \m E_j \m L^{-1}_{\v t}\m T + \m T \frac{\partial\left[\m L^{-1}_{\v t}\m T\right]}{\partial t_j}.
\end{equation}
Furthermore, 
\begin{equation}
    \label{ap:3}
    \frac{\partial\left[\m L^{-1}_{\v t}\m T\right]}{\partial t_j} = \frac{\partial\left[\m L^{-1}_{\v t}\right]}{\partial t_j}\m T+\m L^{-1}_{\v t}\m E_j,
\end{equation}
and using the derivative of an invertible matrix we have,
\begin{equation}
    \label{ap:4}
    \frac{\partial\left[\m L^{-1}_{\v t}\right]}{\partial t_j} = -\m L^{-1}_{\v t}\frac{\partial\m L_{\v t}}{\partial t_j}\m L^{-1}_{\v t},
\end{equation}
and,
\begin{equation}
    \label{ap:5}
    \frac{\partial\m L_{\v t}}{\partial t_j} = \m E_j \m Z\m T+\m T\m Z\m E_j.
\end{equation}

Substituting $\eqref{ap:5}\to\eqref{ap:4}\to\eqref{ap:3}\to\eqref{ap:2}\to\eqref{ap:1}$ we obtain the expression,
\begin{equation}
\label{ap: 6}
  \frac{\partial}{\partial t_j}\left[-\tr\left(\m X^{\top}\widetilde{\m P}(\v t)\m X\right)\right] = -\tr\left[\m K \m E_j\m L^{-1}_{\v t}\m T\m K + \m K\m T \m L^{-1}_{\v t}\m E_j\m K - \m K \m T\m L^{-1}_{\v t}\m E_j\m Z  \m T\m L^{-1}_{\v t}\m T\m K- \m K \m T\m L^{-1}_{\v t}\m T\m Z \m E_j \m L^{-1}_{\v t}\m T\m K\right].
\end{equation}
In order to simplify this expression we consider the following fact. If we have the matrices $\m A= (a_{ij})\in \bb{R}^{n\times n}$ and $\m B=(b_{ij})\in \bb{R}^{n\times n}$ then, $\tr\left(\m A\m E_j \m B\right)=\left(\m B\m A\right)_{jj} = \sum_{i=1}^n a_{ij}b_{ji}$. Using this, we obtain,

\begin{align*}
 \frac{\partial}{\partial t_j}\left[-\tr\left(\m X^{\top}\widetilde{\m P}(\v t)\m X\right)\right] &= -\left[\m L^{-1}_{\v t}\m T\m K^2\right]_{jj} - \left[\m K^2\m T \m L^{-1}_{\v t}\right]_{jj} + \left[\m Z  \m T\m L^{-1}_{\v t}\m T\m K^2 \m T\m L^{-1}_{\v t}\right]_{jj}+ \left[\m L^{-1}_{\v t}\m T\m K^2 \m T\m L^{-1}_{\v t}\m Z\right]_{jj}\\
 & = 2\left[\m L^{-1}_{\v t}\m T\m K^2\left( \m T\m L^{-1}_{\v t}\m Z-\m I\right)\right]_{jj},
\end{align*}
since the matrices $\m L^{-1}_{\v t}$, $\m K$, $\m Z$ and $\m T$ are all symmetric. Considering the partial derivative of the penalty term is $\frac{\partial}{\partial t_j}\left[\lambda\sum_i t_i\right] = \lambda$ we have the following expression for the gradient vector of $f_{\lambda}(\v t)$,
\begin{equation*}
    \nabla f_{\lambda}(\v t) = 2\operatorname{Diag}\left[\m L_{\v t}^{-1}\m T \m K^2 \left(\m T\m L_{\v t}^{-1}\m T\m Z - \m I\right)\right]+\lambda\v 1.
\end{equation*}$\square$

\emph{Proof of \cref{lem: cnr_k}}. For reasons outlined in the proof of \cref{lem: cnr_p} we assume without loss of generality that all the zero-elements in $\v s \in \{0,1\}^n$ appear at the end, in the form,
\[
\v s = (s_1,\dots\,s_l,0,\dots, 0).
\]
where $l$ is equal to the number of non-zeros in $\v s \in \{0,1\}^n$. Then, $\widetilde{\m K}(\v s)$ is given by the block-wise matrix,

\begin{equation}
\label{ps_2}
\widetilde{\m P}(\v s) = \begin{bmatrix} \m K_{[\v s]}  & \m 0\end{bmatrix}\begin{bmatrix} \m  K_{[\v s, \v s]}  & \m 0\\ \m 0 & \delta \m I\end{bmatrix}^{\dagger}\begin{bmatrix} \m K_{[\v s]}^{\top}  \\ \m 0\end{bmatrix}.
\end{equation}
Using the block-diagonal pseudo-inverse formula we have,
\begin{align*}
\widetilde{\m P}(\v s) &= \begin{bmatrix}\m K_{[\v s]}  & \m 0\end{bmatrix}\begin{bmatrix} \m  K_{[\v s, \v s]}^{\dagger}  & \m 0\\ \m 0 & \delta^{-1} \m I\end{bmatrix}\begin{bmatrix} \m K_{[\v s]}^{\top}  \\ \m 0\end{bmatrix}\\
&= \m K_{[\v s]}\m  K_{[\v s, \v s]}^{\dagger} \m K_{[\v s]}^{\top}.
\end{align*}$\square$

\emph{Proof of \cref{lem: cont_k}}. For any positive semi-definite kernel matrix $\m K\in \bb{R}^{n \times n}$, a decomposition of the form  $\m K = \m X^{\top}\m X$ where $ \m X \in \bb{R}^{n \times n}$ always exists. Therefore, the function $\widetilde{\m K}({\v t})$ can be written as,
\begin{align*}
    \widetilde{\m K}({\v t}) &= \m X^{\top}\m X\m T \left[\m T\m X^{\top}\m X\m T+\delta(\m I-\m T^2)\right]^{\dagger}\m T\m X^\top\m X \\
    &= \m X^{\top}\widetilde{\m P}(\v t)\m X^\top.
\end{align*}
From \cref{lem: cont_p} we know that the function $\widetilde{\m P}(\v t)$ is continuous over $[0,1]^n$. Since $\m X$ is not a function of $\v t$ we conclude that $\widetilde{\m K}({\v t})$ is also continuous over $[0, 1]^n$. $\square$

\emph{Proof of \cref{lem: g_grad}}. To obtain the gradient $g_{\lambda}(\v t)$ for $\v t \in (0,1)^n$ we first simplify the term $\| \widetilde{\m K}({\v t})-\m K \|_F^2$. Since $\widetilde{\m K}({\v t})$ and $\m K$ are symmetric, we have the expansion,
\begin{align*}
    \| \widetilde{\m K}({\v t})-\m K \|_F^2 &= \tr\left[\left(\widetilde{\m K}({\v t})-\m K\right)^2\right]\\
    & = \tr\left[\left(\widetilde{\m K}({\v t})\right)^2-2\widetilde{\m K}({\v t}) \m K + \m K^2\right].
\end{align*}
Therefore,

\begin{align*}
    \frac{\partial}{\partial t_j}\left[\| \widetilde{\m K}({\v t})-\m K \|_F^2\right] &= \tr\left( \frac{\partial}{\partial t_j}\left[\left(\widetilde{\m K}({\v t})\right)^2\right]\right)- \tr\left(\frac{\partial}{\partial t_j}\left[2\widetilde{\m K}({\v t}) \m K\right]\right)\\
    &= 2\tr\left(\frac{\partial\widetilde{\m K}({\v t})}{\partial t_j}\widetilde{\m K}({\v t})\right) - 2\tr\left(\frac{\partial\widetilde{\m K}({\v t})}{\partial t_j}\m K\right)\\
    &=  2\tr\left(\frac{\partial\widetilde{\m K}({\v t})}{\partial t_j}\m D\right)
\end{align*}
where $\m D = \widetilde{\m K}({\v t})-\m K$. Factorize $\m K$ as $\m K = \m X^{\top}\m X$ and let $\m Z =\m K - \delta \m I$ and $\m L_{\v t} = \m T \m Z \m T +\v\delta\m I$. Then, notice that the derivative $\frac{\partial\widetilde{\m K}({\v t})}{\partial t_j} = \frac{\partial\left[\m X^{\top}\widetilde{\m P}(\v t)\m X\right]}{\partial t_j} $ is the same expression that we derived in the proof of \cref{lemma: f_grad}, see \eqref{ap: 6}. Substituting this expression in for $\frac{\partial\widetilde{\m K}({\v t})}{\partial t_j}$, we obtain,
\begin{equation*}
    \frac{\partial}{\partial t_j}\left[\| \widetilde{\m K}({\v t})-\m K \|_F^2\right] = 2\tr\left[\m K \m E_j\m L^{-1}_{\v t}\m T\m K\m D + \m K\m T \m L^{-1}_{\v t}\m E_j\m K\m D - \m K \m T\m L^{-1}_{\v t}\m E_j\m Z  \m T\m L^{-1}_{\v t}\m T\m K\m D- \m K \m T\m L^{-1}_{\v t}\m T\m Z \m E_j \m L^{-1}_{\v t}\m T\m K\m D\right].
\end{equation*}
Once again, using the fact that $\tr\left(\m A\m E_j \m B\right)=\left(\m B\m A\right)_{jj} = \sum_{i=1}^n a_{ij}b_{ji}$ for matrices $\m A= (a_{ij})\in \bb{R}^{n\times n}$ and $\m B=(b_{ij})\in \bb{R}^{n\times n}$,  we obtain,
\begin{align}
    \label{ap: 7}
    \frac{\partial}{\partial t_j}\left[\| \widetilde{\m K}({\v t})-\m K \|_F^2\right] &= 2\left[\left(\m L^{-1}_{\v t}\m T\m K\m D\m K\right)_{jj}+\left(\m K\m D\m K\m T \m L^{-1}_{\v t}\right)_{jj}-\left(\m Z  \m T\m L^{-1}_{\v t}\m T\m K\m D\m K \m T\m L^{-1}_{\v t}\right)_{jj}-\left(\m L^{-1}_{\v t}\m T\m K\m D\m K \m T\m L^{-1}_{\v t}\m T\m Z\right)_{jj}\right]\\
    & = 4\left[\left(\m L^{-1}_{\v t}\m T\m K\m D\m K\right)_{jj}-\left(\m L^{-1}_{\v t}\m T\m K\m D\m K \m T\m L^{-1}_{\v t}\m T\m Z\right)_{jj}\right],
\end{align}
since the matrices $\m L^{-1}_{\v t}$, $\m K$, $\m Z$ and $\m T$ are all symmetric. Considering the partial derivative of the penalty term is $\frac{\partial}{\partial t_j}\left[\lambda\sum_i t_i\right] = \lambda$ we have the following expression for the gradient vector of $g_{\lambda}(\v t)$,
\begin{equation*}
    \nabla g_{\lambda}(\v t) = 4\operatorname{Diag}\left[\m L_{\v t}^{-1}\m T \m K\m D\m K \left( \m I  -\m T\m L_{\v t}^{-1}\m T\m Z\right)\right]+\lambda\v 1.
\end{equation*}$\square$
\emph{Proof of \cref{lemma: stoch_f}}. From \cref{lemma: f_grad}, we have, 
\begin{align*} 
    \nabla f_{\lambda}(\v t) &= 2\operatorname{Diag}\left[\m L_{\v t}^{-1}\m T \m K^2 \left(\m T\m L_{\v t}^{-1}\m T\m Z - \m I\right)\right]+\lambda\v 1\\
    &= 2\left[\operatorname{Diag}\left(\m L_{\v t}^{-1}\m T \m K^2 \m T\m L_{\v t}^{-1}\m T\m Z\right) -\operatorname{Diag}\left(\m L_{\v t}^{-1}\m T \m K^2 \right)\right]+\lambda\v 1\\
    &= 2\v \alpha+\lambda\v 1,
\end{align*}
where $\v \alpha = \operatorname{Diag}\left(\m L_{\v t}^{-1}\m T \m K^2 \m T\m L_{\v t}^{-1}\m T\m Z\right) -\operatorname{Diag}\left(\m L_{\v t}^{-1}\m T \m K^2 \right)$. To obtain an unbiased estimator for $\v \alpha$, we use the factorized estimator $\hat{\ell}$ for the diagonal of a square matrix. Recall, to estimate the diagonal of the matrix $\m A = \m B \m C^{\top}$ where $\m A, \m B, \m C \in \bb{R}^{n\times n}$ we let $\v z \in \bb{R}^n$ be a random vector sampled from the Rademacher distribution. Then the unbiased estimate for $\operatorname{Diag}\left(\m A\right)$ is $\hat{\ell}= \m B\v z\odot\m C\v z$ (see \citep{martens2012estimating, mathur2021variance} for proof and analysis). We factorize the matrices in $\v \alpha$ so that,
\[
\v \alpha = \operatorname{Diag}\left(\overbrace{\m L_{\v t}^{-1}\m T \m K }^{\m B_1}\overbrace{\m K \m T\m L_{\v t}^{-1}\m T\m Z}^{\m C_1^{\top}}\right) -\operatorname{Diag}\left(\overbrace{\m L_{\v t}^{-1}\m T \m K}^{\m B_2}\overbrace{\m K}^{\m C_2^{\top}} \right)
\]
Then, the factorized estimator for $\v \alpha$ is given by,
\[
\v \phi = \m L_{\v t}^{-1}\m T \m K \v z \odot \m Z\m T\m L_{\v t}^{-1}\m T\m K\v z -  \m L_{\v t}^{-1}\m T \m K \v z\odot \m K \v z,
\]
where $\bb{E}[\v \phi] = \v \alpha$ and $\v z \in \bb{R}^n$ is a Rademacher random variable. If we compute the following variables: $(1)\,\v a= \m K\v z,\,(2)\, \v b = \m L_{\v t}^{-1}( \v t\odot \v a)$, we have,
\[
   (1) \, \v a = \m K \v z,\quad\v (2) \, \v b = \m L_{\v t}^{-1}\m T \m K \v z
\] 
and $\v \phi$ simplifies to,
\[
\v \phi = \v b \odot \m Z(\v t \odot \v b) - \v a \odot \v b.
\]
Therefore, for $\v t \in (0,1)^n$ we have,
\[\nabla f_{\lambda}(\v t) = 2\bb{E}\left[\v \phi\right]+\lambda \m 1.\]
$\square$

\emph{Proof of \cref{lemma: stoch_g}}. From \cref{lem: g_grad} and \cref{ap: 7}, we have, 
\begin{align*} 
    \nabla g_{\lambda}(\v t) &=4\operatorname{Diag}\left[\m L_{\v t}^{-1}\m T \m K\m D\m K \left( \m I  -\m T\m L_{\v t}^{-1}\m T\m Z\right)\right]+\lambda\v 1.\\
    &= 2\left[\operatorname{Diag}\left(\m L_{\v t}^{-1}\m T \m K\m D\m K \right) +\operatorname{Diag}\left( \m K\m D\m K\m T\m L_{\v t}^{-1}\right) -\operatorname{Diag}\left(\m Z\m T\m L_{\v t}^{-1}\m T \m K\m D\m K\m T \m L_{\v t}^{-1}\right)-\operatorname{Diag}\left(\m L_{\v t}^{-1}\m T \m K\m D\m K \m T\m L_{\v t}^{-1}\m T\m Z\right) \right]+\lambda\v 1\\
    &= 2\v \beta+\lambda\v 1.
\end{align*}
 To obtain an unbiased estimator for $\v \beta$, we once again use the factorized estimator for the diagonal of a square matrix. We factorize the matrices in $\v \beta$ so that,
\[
\v \beta = \operatorname{Diag}\left(\overbrace{\m L_{\v t}^{-1}\m T \m K\m D}^{\m B_1}\overbrace{\m K}^{\m C_1^{\top}} \right) +\operatorname{Diag}\left(\overbrace{ \m K\m D}^{\m B_2}\overbrace{\m K\m T\m L_{\v t}^{-1}}^{\m C_2^{\top}}\right) -\operatorname{Diag}\left(\overbrace{\m Z\m T\m L_{\v t}^{-1}\m T \m K\m D}^{\m B_3}\overbrace{\m K\m T \m L_{\v t}^{-1}}^{\m C_3^{\top}}\right)-\operatorname{Diag}\left(\overbrace{\m L_{\v t}^{-1}\m T \m K\m D}^{\m B_4}\overbrace{\m K \m T\m L_{\v t}^{-1}\m T\m Z}^{\m C_4}\right).
\]
Then, the factorized estimator for $\v \beta$ is given by,
\[
\v \psi = \m L_{\v t}^{-1}\m T \m K \m D\v z \odot \m K \v z+ \m K \m D \v z \odot \m L_{\v t}^{-1}\m T \m K \v z - \m Z\m T\m L_{\v t}^{-1}\m T\m K \m D\v z \odot \m L_{\v t}^{-1}\m T\m K \v z -  \m L_{\v t}^{-1}\m T \m K \m D \v z\odot \m Z\m T\m L_{\v t}^{-1}\m T\m K \v z.
\]
where $\bb{E}[\v \psi] = \v \beta$ and $\v z \in \bb{R}^n$ is a Rademacher random variable. Recall that $\m D = \widetilde{\m K}({\v t})-\m K$ and $\widetilde{\m K}({\v t}) = \m K \m T  \m L_{\v t}^{-1} \m T \m K$. Then, if we compute the following variables: $(1)\,\v a= \m K \v z,\,(2)\,\v b = \m L_{\v t}^{-1}( \v t \odot \v a),\,(3)\,\v c=\m K(\v t\odot \v b)-\v a,\, (4)\,\v d = \m K \v c,\,(5)\,\v e = \m L_{\v t}^{-1}(\v t \odot \v d)$, we have,
\[
   (1) \, \v a = \m K \v z,\quad\v (2) \, \v b = \m L_{\v t}^{-1}\m T \m K \v z,\quad (3) \, \v c = \m D \v z,\quad (4) \,\v d = \m K \m D \v z\quad (5) \,\v e = \m L_{\v t}^{-1}\m T \m K \m D \v z,
\]
and $\v \psi$ simplifies to,
\[\v \psi = \v b \odot \v d+\v a \odot \v e - \v e\odot \m Z(\v t \odot\v b)-\v b\odot \m Z(\v t \odot\v e).\]
Hence, we have the expression for the gradient,
\[\nabla g_{\lambda}(\v t) = 2\bb{E}\left[\v \psi\right]+\lambda \m 1.\]
$\square$
\emph{Proof of \cref{lem: cont_p} }For the same reasons outlined in the proof of \cref{lem: cnr_p} we assume without loss of generality that all the zero-elements in $\v t \in \{0,1\}^n$ appear at the end, in the form,
\[
\v t = (t_1,\dots\,t_l,0,\dots, 0).
\]
Then $\m L_{\v t}$ is given by,
 \[
 \m L_{\v t} = \begin{bmatrix}  (\m T)_{+}(\m K)_{+}(\m T)_{+}+\delta(\m I - (\m T)^2_{+})  & \m 0\\ \m 0 & \delta \m I\end{bmatrix}  = \begin{bmatrix} (\m L_{\v t})_{+}  & \m 0\\ \m 0 & \delta \m I\end{bmatrix}
 \]
 and since $(\v t)_{+}\in [0,1)^l$ the matrix $ (\m L_{\v t})_{+}$ is invertible. Therefore, 
\begin{align*}
    \m L_{\v t}^{-1}(\v t \odot \v r) &= \begin{bmatrix} (\m L_{\v t})^{-1}_{+}  & \m 0\\ \m 0 & \delta^{-1} \m I\end{bmatrix}\begin{bmatrix} (\v t)_+ \odot (\v r)_+  \\ \m 0\end{bmatrix}\\
    & = \begin{bmatrix}(\m L_{\v t})^{-1}_{+}\left((\v t)_+ \odot (\v r)_+\right)  \\ \m 0\end{bmatrix}.
\end{align*}
$\square$
\end{document}